\documentclass[twocolumn,twocolappendix,times]{aastex631}

\graphicspath{{./}{figures/}}

\usepackage{color} 
\usepackage{amsmath}
\usepackage{float}
\usepackage{amssymb}
\usepackage{graphicx}
\usepackage{natbib}
\usepackage{multirow}
\usepackage{array}
\usepackage{makecell}

\definecolor{emerald}{rgb}{0.1,0.5,0.3}

\definecolor{purple}{HTML}{DA70D6}

\definecolor{gray}{HTML}{A9A9A9}

\shorttitle{AGN in the Post-Starburst Phase}
\shortauthors{Luo et al.}

\begin{document}


\title{A Multiwavelength Evaluation of AGN in the Post-Starburst Phase}


\author[0000-0002-0696-6952]{Yuanze Luo}
\affiliation{William H. Miller III Department of Physics and Astronomy, Johns Hopkins University, Baltimore, MD 21218, USA}
\affiliation{Department of Physics and Astronomy and George P. and Cynthia Woods Mitchell Institute for Fundamental Physics and Astronomy, Texas A\&M University, 4242 TAMU, College Station, TX 77843-4242, US}

\author[0000-0001-7883-8434]{Kate Rowlands}
\affiliation{AURA for ESA, Space Telescope Science Institute, 3700 San Martin Drive, Baltimore, MD 21218, USA}
\affiliation{William H. Miller III Department of Physics and Astronomy, Johns Hopkins University, Baltimore, MD 21218, USA}

\author[0000-0002-4261-2326]{Katherine Alatalo}
\affiliation{Space Telescope Science Institute, 3700 San Martin Dr, Baltimore, MD 21218, USA}
\affiliation{William H. Miller III Department of Physics and Astronomy, Johns Hopkins University, Baltimore, MD 21218, USA}

\author[0000-0002-3249-8224]{Lauranne Lanz}
\affiliation{Department of Physics, The College of New Jersey, Ewing, NJ 08628, USA}

\author[0000-0001-6670-6370]{Timothy Heckman}
\affiliation{William H. Miller III Department of Physics and Astronomy, Johns Hopkins University, Baltimore, MD 21218, USA}
\affiliation{School of Earth and Space Exploration, Arizona State University, Tempe, AZ 85287-1404, USA}

\author[0000-0001-6245-5121]{Elizaveta Sazonova}
\affiliation{Waterloo Centre for Astrophysics, University of Waterloo, Waterloo, ON, N2L 3G1 Canada}
\affiliation{Department of Physics and Astronomy, University of Waterloo, Waterloo, ON N2L 3G1, Canada}

\author[0000-0002-9471-8499]{Pallavi Patil}
\affiliation{William H. Miller III Department of Physics and Astronomy, Johns Hopkins University, Baltimore, MD 21218, USA}


\author[0000-0001-9328-3991]{Omar Almaini}
\affiliation{School of Physics and Astronomy, University of Nottingham, University Park, Nottingham NG7 2RD, U.K.}

\author[0000-0002-1759-6205]{Vincenzo R. D'Onofrio}
\affiliation{Department of Physics and Astronomy and George P. and Cynthia Woods Mitchell Institute for Fundamental Physics and Astronomy, Texas A\&M University, 4242 TAMU, College Station, TX 77843-4242, US}

\author[0000-0002-4235-7337]{K. Decker French}
\affiliation{University of Illinois Urbana-Champaign Department of Astronomy, University of Illinois, 1002 W. Green St., Urbana, IL 61801, USA}

\author[0000-0003-3191-9039]{Justin Otter}
\affiliation{William H. Miller III Department of Physics and Astronomy, Johns Hopkins University, Baltimore, MD 21218, USA}

\author[0000-0003-4030-3455]{Andreea O. Petric}
\affiliation{William H. Miller III Department of Physics and Astronomy, Johns Hopkins University, Baltimore, MD 21218, USA}
\affiliation{Space Telescope Science Institute, 3700 San Martin Dr, Baltimore, MD 21218, USA}

\author[0000-0002-4430-8846]{Namrata Roy}
\affiliation{Center for Astrophysical Sciences, Department of Physics and Astronomy, Johns Hopkins University, Baltimore, MD, 21218}

\author[0009-0004-0844-0657]{Maya Skarbinski}
\affiliation{William H. Miller III Department of Physics and Astronomy, Johns Hopkins University, Baltimore, MD 21218, USA}

\author[0000-0003-3256-5615]{Justin S. Spilker}
\affiliation{Department of Physics and Astronomy and George P. and Cynthia Woods Mitchell Institute for Fundamental Physics and Astronomy, Texas A\&M University, 4242 TAMU, College Station, TX 77843-4242, US}

\author[0000-0003-1535-4277]{Margaret E. Verrico}
\affiliation{University of Illinois Urbana-Champaign Department of Astronomy, University of Illinois, 1002 W. Green St., Urbana, IL 61801, USA}
\affiliation{Center for AstroPhysical Surveys, National Center for Supercomputing Applications, 1205 West Clark Street, Urbana, IL 61801, USA}

\author[0000-0002-8956-7024]{Vivienne Wild}
\affiliation{School of Physics and Astronomy, University of St Andrews, North Haugh, St Andrews, KY16 9SS, U.K.}
\begin{abstract}

The quenching of star formation is a crucial phase in galaxy evolution. Although active galactic nuclei (AGN) feedback has been proposed as a key driver of this transition, the lack of strong AGN in nearby quenching galaxies raises questions about its effectiveness. In this study, we investigate AGN activity in post-starburst galaxies (PSBs), star-forming galaxies (SFGs), and quiescent galaxies (QGs) at $z<$ 0.2, using multiwavelength data from eROSITA/eFEDS (X-ray), WISE (mid-infrared), and FIRST (radio). We assess AGN incidence and strength across different stages and apply stacking techniques to undetected galaxies to recover average AGN properties. Comparisons between observed luminosity and that expected from star formation (L$_{\rm obs}$/L$_{\rm SF}$) show that PSBs are consistent with star formation dominating their radio and X-ray emission. Although PSBs exhibit a MIR AGN incidence rate twice that of SFGs, their estimated AGN luminosities are small compared to those of MIR AGN in the literature. PSBs overall do not display significantly enhanced AGN emission relative to mass- and redshift-matched SFGs and QGs. While the presence of obscured, low-luminosity AGN in PSBs cannot be excluded, such AGN, if present, could be fueled by residual gas from the preceding starburst and may not play a dominant role in quenching. Our findings suggest that AGN's role in quenching at low redshift is more subtle than violently removing the gas -- the feedback is likely more ``preventive" than ``ejective”.

\end{abstract}

\section{Introduction} \label{sec:intro}

Galaxies up to $z\sim3$ exhibit a well-established bimodality in both morphology and color \citep{strat2001,Baldry_2004,Willmer_2006,Brammer_2009,Jin_2014}, with the majority in either the ``blue cloud" (spiral, star-forming galaxies) or the ``red sequence" (quiescent, elliptical galaxies). Observations show a steady increase in both the number density and total stellar mass of quiescent galaxies towards the present day \citep{Ilbert_2013,Muzzin_2013}, prompting an evolutionary scenario where galaxies evolve from blue to red \citep[e.g.,][]{Bell_2004,Faber_2007,Weaver_2023,Clausen_2024}. This transformation happens as star formation (SF) in the galaxy dwindles and the decline in star formation rate (SFR) is referred to as ``quenching”.

Active galactic nuclei (AGN) have been invoked as an important mechanism for quenching, as cosmological simulations predict that they are capable of driving powerful outflows to expel gas from the galaxy \citep[e.g.,][]{hopkins2008} and offset high cooling rates in galaxy halos to prevent gas from falling in \citep[e.g.,][]{Weinberger_2017}. However, in the local universe ($z\lesssim$ 0.5), observational evidence of galaxy quenching from AGN feedback remains circumstantial \citep[e.g.,][]{Smethurst_2017,French_2018}. Large molecular gas reservoirs have been found in quenched galaxies \citep[e.g.,][]{Rowlands_2015,French_2015,Otter_2022}, suggesting that galaxies do not cease SF solely by depleting their gas reservoirs and SF suppression could be due to turbulent heating within the galaxy \citep[e.g.,][]{Smercina_2022,Brunetti_2024}. Furthermore, AGN identified in nearby rapidly quenching/quenched galaxies tend to be weak (e.g., \citealt{Roy_2021_wind,Roy_2021_morph,Lanz_2022,luo2022,French_2023}; similarly at cosmic noon, see \citealt{Almaini_2025}), raising questions about the efficacy of AGN in quenching and its relative importance compared to other mechanisms.

Post-starburst galaxies (PSBs) have quenched both recently and rapidly within the last $\sim$1 Gyr \citep[e.g.,][]{Quintero_2004,Goto_2007,Snyder_2011}. They probe the intermediate stage between star-forming and quiescent and are ideal laboratories to study quenching mechanisms because they likely retain signatures of their quenching triggers \citep[e.g.,][]{Yang_2004,Yang_2008}. Studies have reported preliminary evidence for a slight enhancement in high-ionization optical emission lines \citep{Yan_2006,Wild_2007,Wild_2010}, X-ray \citep{Brown_2009,Lanz_2022}, mid-infrared (MIR, \citealt{Alatalo_2017}) and radio emission \citep{Nielsen_2012,Meusinger_2017} in some low-redshift PSBs, hinting at an over-abundance of AGN in the PSB phase. This over-abundance could be a sign of AGN playing a role in quenching or enhanced fueling of AGN following starbursts (SBs). However, previous studies have often been limited in sample sizes and to single-wavelength analyses, hindering our ability to obtain a comprehensive view of AGN prevalence at different galaxy evolutionary stages in the local universe. In this study, we qualitatively assess AGN dominance utilizing a large sample of local PSBs, compiled from multiple sources, through a multiwavelength approach using X-ray, MIR, and radio data.


Different wavelengths provide complementary insights into AGN activity (see \citealt{Padovani_2017} for a review). X-ray observations provide a robust way of identifying AGN because X-ray emission from other astrophysical processes (e.g., SF) is typically weak by comparison \citep{Brandt_2015}. X-ray emission from AGN is predominantly produced by the inverse Compton scattering of photons from the accretion disk \citep[e.g.,][]{Hickox_2018}. Thus X-ray emission directly probes active accretion onto the black hole and is often associated with radiatively efficient accretion at Eddington ratios $\gtrsim$ 0.01 \citep[e.g.,][]{Shakura_1973,Fabian_2012}. MIR emission from AGN originates from hot dust heated by the X-ray flux from the accretion disk corona re-radiating at longer wavelengths \citep[e.g.,][]{Lacy_2004,Stern_2005}. MIR emission is therefore particularly important in identifying obscured AGN, when X-ray and optical emission is hard to observe. Radio emission traces the synchrotron radiation from AGN-driven jets and is often associated with the ``radio/kinetic" mode of AGN feedback \citep[e.g.,][]{Croton_2006,Somerville_2008}. Radio emission is not susceptible to extinction, can help pinpoint both obscured and unobscured AGN, and provides key insights in estimating the kinetic power from AGN feedback \citep[e.g.,][]{Heckman_2024}. As different wavelengths trace different components of AGN, synergistic analyses of multiwavelength data on a statistical sample of PSBs could help build a census of AGN in the post-starburst population. In the low-luminosity regime, stacking has proven to be a powerful technique for uncovering the average properties of AGN populations \citep[e.g.,][]{Toba_2022,Ito_2022}. By applying a consistent methodology to nearby star-forming galaxies (SFGs) and quiescent galaxies (QGs), this study will help determine whether black hole growth, as traced by AGN activity, is most prevalent before, during, or after the quenching phase. We note that this work focuses on the nearby universe ($z<$ 0.2) and the low end of the AGN luminosity function. At higher redshifts and AGN luminosities, powerful AGN-driven radio jets could be responsible for most of the quenching over cosmic time \citep[e.g.,][]{Heckman_2024}.


This paper is organized as follows. In Section \ref{sec:data} we present the sample selection of SFGs, PSBs, and QGs, along with the multiwavelength data used in this study. In Section \ref{sec:analysis} we describe our stacking analysis at each wavelength and corresponding results. We further investigate evidence for AGN across different galaxy populations at each wavelength in Sections \ref{sec:agn in radio}, \ref{sec: agn in mir}, and \ref{sec: agn in xray}, and discuss possible implications and subtleties on the picture of AGN quenching in the local universe in Section \ref{sec:discussion}. We summarize our findings in Section \ref{sec: conclusions}. Throughout this paper, we assume a flat cosmological model with $H_0=70\ \rm{km\ s^{-1}\ Mpc^{-1}}$, $\Omega_m=0.3$, and $\Omega_{\Lambda}=0.7$. 





\section{Sample and Data} \label{sec:data}

\begin{figure*}
\centering
\includegraphics[width=\textwidth]{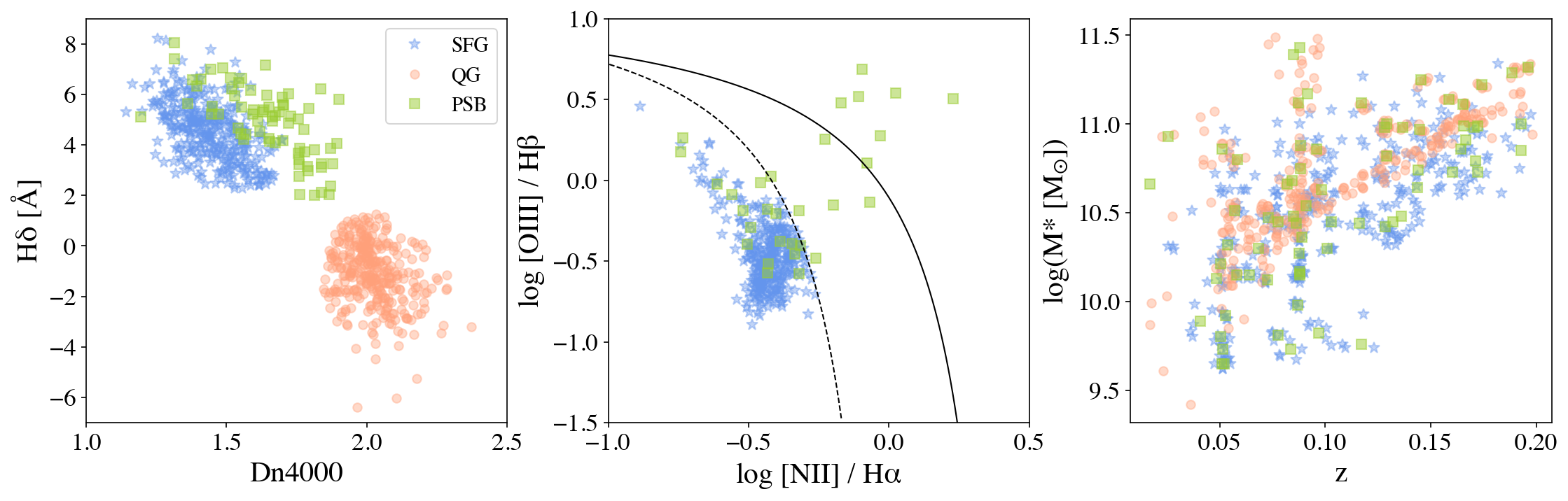}
\caption{\textbf{Left:} The distribution of SFGs, PSBs and QGs on the H$\delta_A$ vs. Dn4000 plot. \textbf{Middle:} [\ion{O}{3}]/H$\beta$ vs. [\ion{N}{2}]/H$\alpha$ BPT diagram. The dashed and solid lines are demarcation lines for SF and AGN ionization, from \citet{Kauffmann_2003} and \citet{Kewley_2001}, respectively. Only galaxies with SNR $\geqslant$ 3 for all four emission lines are shown. QGs are selected to have minimal emission (Section \ref{sec:control sample}) and therefore not plotted here. \textbf{Right:} Stellar mass vs. redshift for the PSB sample and mass- and redshift-matched comparison samples of SFGs and QGs. The vertical and diagonal aggregations of QGs in this panel are not biases introduced by the matching process. See Section \ref{sec:control sample} and Appendix \ref{appendix control sample} for more details.
\label{fig:sample}}
\end{figure*}

We utilize data from the eROSITA Final Equatorial Depth Survey (eFEDS, \citealt{Brunner_2022}) for the X-ray analysis and the Faint Images of the Radio Sky at Twenty centimeters (FIRST; \citealt{Becker_1995}) for the radio analysis. The MIR data is from the Wide-field Infrared Survey Explorer (WISE; \citealt{Wright_2010}). We select galaxies covered by all three facilities for multiwavelength analyses as explained in Section \ref{sec:galaxy sample}. Specific data products used for each survey are described in Section \ref{sec:surveys}.

\subsection{Samples of PSBs, SFGs, and QGs from SDSS} \label{sec:galaxy sample}

Our sample of galaxies is classified spectroscopically based on data from the Sloan Digital Sky Survey (SDSS), mostly Data Release 7 (DR7, \citealt{Abazajian_2009}). We obtain their redshifts from the SDSS database, D$_{\rm n}$4000\footnote{Defined as the ratio of the average flux density F$_\nu$ in the bands 3850--3950 and 4000--4100\,\AA\ as in \citet{Balogh_1999}.} from the MPA-JHU catalog \citep{Kauffmann_2003-mpajhu,Brinchmann_2004,Tremonti_2004}, and stellar mass (M$_*$) from the spectral energy distribution (SED) fitting of SDSS and WISE data in \citet{Chang_2015}\footnote{\url{https://irfu.cea.fr/Pisp/yu-yen.chang/sw.html}}. We obtain absorption and emission line measurements for our galaxies from the OSSY catalog \citep{Oh_2011} based on SDSS DR7. For the small fraction ($\lesssim$ 5\%) of galaxies in OSSY without M$_*$ in \citet{Chang_2015}, we use M$_*$ from the MPA-JHU catalog\footnote{M$_*$ in these two catalogs are consistent. See Section 4.1 in \citet{Chang_2015} for more discussion.}. 

\subsubsection{PSB Sample} \label{sec:psb sample}

The bursty star formation history of PSBs leaves special imprints in their optical spectra: PSBs exhibit deep Balmer absorption lines associated with A-type stars that have a lifetime $\sim$ 1 Gyr. Spectra with such strong Balmer absorption lines and weaker than average emission lines are thus characteristic of a recent ($\lesssim$ 1 Gyr) starburst event followed by a decline in SF. Our parent PSB sample contains over 6000 PSBs at $z<$ 0.2, which comes from a compilation of published catalogs or methods (\citealt{Wild_2007,Goto_2007,Alatalo_2016,Pattarakijwanich_2016,Wild_2025}; Tremonti et al. in prep). Specific PSB criteria and notes on the samples are summarized in Table 1 of \citet{Wild_2025}. We briefly outline the PSB criteria in each literature sample that comprises our parent PSB sample below, where positive line equivalent width (EW) corresponds to absorption:
    \begin{itemize}
    \item \citet{Goto_2007}: EW(H$\delta$) $>$ 5\AA, EW(H$\alpha$) $>$ $-$3\AA, EW([\ion{O}{2}]) $>$ $-$2.5\AA; SDSS DR5
    \item \citet{Wild_2007,Wild_2025}: Principal component (PC) constraints: PC2 $>$ 0.025 and PC1 $<$ $-$1.5; additional cut on Balmer decrement (H$\alpha$/H$\beta>$ 5.2 and
    H$\alpha$/H$\beta>$ 6.6 in the low- and high-mass regime, respectively) following \citet{Pawlik_2018} to ensure the reliability of the PC method; utilizing the method in \citet{Wild_2007}; the PC cuts follow \citet{Wild_2025} to select excess Balmer absorption compared to 4000\AA\ break strength; SDSS DR7\footnote{\url{http://star-www.st-andrews.ac.uk/~vw8/downloads/DR7PCA.html}}
    \item \citet{Alatalo_2016}: EW(H$\delta$) $>$ 5\AA; emission line ratios consistent with shock ionization and inconsistent with SF ionization; the Shocked Post-starburst Galaxy Survey (SPOGS); SDSS DR7
    \item \citet{Pattarakijwanich_2016}: EW(H$\delta$) $>$ 4\AA, EW([\ion{O}{2}]) $>$ $-$2.5\AA, A/(A + K) $>$ 0.25 from the template fitting method in \citet{Quintero_2004}; cover 0.05 $\lesssim z \lesssim$ 1.3, but we include only $z <$ 0.2 PSBs; SDSS DR9
    \item Tremonti et al. in prep: H$\delta_{\rm A}$\footnote{Lick Index \citet{Worthey_1994,Worthey_1997}.} $>$ 3\AA, EW(H$\alpha$) $>$ $-$10\AA, log$_{10}$($|$EW(H$\alpha$)$|$) $<$ 0.23$\times$ H$\delta_{\rm A}$ $-$ 0.46; utilizing the method in \citet{Chen_2019}; SDSS DR7
\end{itemize}

In addition to strong Balmer absorption, some selection criteria (\citealt{Goto_2007,Pattarakijwanich_2016,Chen_2019}; Tremonti et al. in prep) require weak emission lines, while others (\citealt{Wild_2007,Alatalo_2016}) do not select against emission lines. PSBs selected without emission line strength cuts typically have younger post-burst age ($\lesssim400$ Myr, e.g., \citealt{French_2018}). Here we combine PSBs from a variety of criteria to ensure a broad coverage of post-burst ages.

The multiwavelength PSB sample consists of galaxies in the parent PSB sample covered by the three aforementioned surveys (eFEDS, WISE, FIRST). We additionally require M$_*>$10$^{9.5}$ M$_{\odot}$ to avoid contamination by dwarf galaxies and to ensure a reasonable match with the comparison samples (Section \ref{sec:control sample}). The final multiwavelength PSB sample contains 73 PSBs. Of the final 73 PSBs in our sample, 53 are selected based on the criteria from \citet{Wild_2007}, with some overlap among the other PSB samples described above. Additionally, 8 PSBs are drawn from the \citet{Pattarakijwanich_2016} sample, 8 from the \citet{Alatalo_2016} sample, and 4 from the Tremonti et al. (in prep) sample (one of which also in \citealt{Goto_2007}).

\subsubsection{Comparison Samples of SFGs and QGs} \label{sec:control sample}
We construct comparison samples of SFGs and QGs matched in both redshift and stellar mass to the PSBs. These samples broadly represent different galaxy evolutionary stages, but we do not imply direct evolutionary links among our specific selections. We utilize the OSSY catalog \citep{Oh_2011}, which provides absorption and emission line measurements based on SDSS DR7. We select galaxies with good observation and fitting quality by requiring \texttt{S\_SN}\footnote{Signal/statistical noise in the OSSY catalog} $\geqslant$ 10 and \texttt{NSIGMA}\footnote{A measurement of the continuum fitting quality indicating the deviation from the median distribution in the OSSY catalog.} $<$ 3. We also require galaxies to have measured properties within reasonable ranges: 0.5 $\leqslant$ Dn4000 $\leqslant$ 3, $-7$ $\leqslant$ H$\delta_A$ [\AA] $\leqslant$ 10, and log(M$_*$ [M$_{\odot}$]) $\geqslant$ 5.

We separate galaxies satisfying the above quality cuts into SFGs and QGs using both the Baldwin, Phillips, and Terlevich diagram \citep[BPT;][]{Baldwin_1981} and H$\delta_A$ vs.\,Dn4000 diagrams \citep[e.g.,][]{Kauffmann_2003}. We fit galaxies on the H$\delta_A$ vs.\,Dn4000 plane with a Gaussian mixture model consisting of three components, representing SFGs, QGs, and others. To investigate and compare multiwavelength properties between PSBs and uncontaminated SFGs and QGs, we further require SFGs to lie in the SF region (below the \citealt{Kauffmann_2003} demarcation line) on the [\ion{O}{3}]/H$\beta$ vs [\ion{N}{2}]/H$\alpha$ BPT diagram with SNR $\geqslant$ 3 for all four emission lines, and QGs to have SNR $<$ 1 for all four emission lines. We exclude from the SFGs and QGs any PSBs in the parent PSB sample. The above quality and emission line requirements on the OSSY catalog leave us with more than 10$^5$ SFGs and QGs, respectively. To ensure the same multiwavelength coverage as the PSB sample, we keep only SFGs and QGs covered by the same surveys (eFEDS, WISE, FIRST). The final parent comparison samples contain 1136 SFGs and 1102 QGs, which are sufficient for our matching algorithm below.


We assign 5 distinct SFGs and QGs, respectively, to each PSB in our multiwavelength PSB sample, following the ``Hungarian algorithm" in \citet{Sazonova_2021}. This algorithm minimizes the total difference in M$_*$ and redshift between the PSB and all its matches (see Appendix \ref{appendix control sample} for the matching schematic plot). We further require the M$_*$ difference between the PSB and its corresponding SFGs and QGs to be less than 0.4 dex\footnote{The typical 1$\sigma$ uncertainty in M$_*$ is $\sim$0.15 dex in the \citet{Chang_2015} and MPA-JHU catalogs used here. To balance having a sufficient number of matched galaxies and maintaining close matches, we adopt the 0.4 dex threshold, which retains roughly 90\% of the originally matched QGs while keeping the mass offset $\lesssim2.5\sigma$. $\Delta$M$_*$ for SFGs are generally smaller due to more available low mass SFGs at $z<0.2$, resulting slightly more SFGs than QGs.} to ensure a close match in M$_*$. The differences in $z$ between PSBs and corresponding comparison galaxies are mostly $\lesssim$ 0.01. We end up with 73 PSBs, 346 SFGs, 333 QGs, and their locations on diagnostic diagrams are shown in Figure \ref{fig:sample}. Although there appear to be some patterns in the distribution of QGs in the M$_*$ vs.\,$z$ panel, they are not biases introduced by the matching process. See Appendix \ref{appendix control sample} for more details.

While the clean baseline samples of SFGs and QGs constructed above facilitate a clear characterization of the average multiwavelength properties for each galaxy population, minimizing the risk of AGN dominating the stacked signals, they may introduce biases when comparing AGN prevalence (e.g., in the MIR, Section \ref{sec: agn in mir}). We show in Appendix \ref{appendix new sample} that alternative comparison samples without the removal of BPT AGN produce the same results on MIR AGN incidence rate as in Section \ref{sec: agn in mir}. Six of our PSBs fall in the AGN region of the BPT diagram because some PSB criteria do not select against emission lines (Section \ref{sec:psb sample}). These BPT-AGN PSBs are included in our subsequent analyses, as their presence does not affect our conclusions: we do not find bright AGN or excess AGN emission relative to SFGs and QGs in PSBs, even if a few optical AGN are present in the PSB sample. Additional details are provided in the relevant sections below and Appendix \ref{appendix new sample}.

\begin{table*}
\centering
\caption{Detection rates at different wavelengths for our samples. Counting errors following Poisson statistics are included for reference (N$^{+\sqrt{N+0.75}+1}_{-\sqrt{N-0.25}}$ for N $\lesssim$ 10; N$^{+\sqrt{N}}_{-\sqrt{N}}$ for N $>$ 10, see \citealt{Gehrels_1986}).} 
\label{tab:det_rate}
\setlength{\tabcolsep}{8pt} 
\renewcommand{\arraystretch}{1.3} 
\begin{tabular}{@{}lccc@{}}
\hline \hline
& \textbf{SFG} & \textbf{PSB} & \textbf{QG} \\
\hline
X-ray (eFEDS) & 7$^{+4}_{-3}$ / 346 (2.0$^{+1.2}_{-0.9}$\%) & 4$^{+3}_{-2}$ / 73 (5.5$^{+4.1}_{-2.7}$\%) & 8$^{+4}_{-3}$ / 333(2.4$^{+1.2}_{-0.9}$\%)\\
MIR - W1 & 340 / 340 (100\%) & 72 / 72 (100\%) & 332 / 332 (100\%) \\
MIR - W2 & 340 / 340 (100\%) & 72 / 72 (100\%) & 332 / 332 (100\%) \\
MIR - W3 & 340 / 340 (100\%) & 63$^{+8}_{-8}$ / 72 (87.5$^{+11.1}_{-11.1}$\%) & 89$^{+9}_{-9}$ / 332 (26.8$^{+2.7}_{-2.7}$\%) \\
MIR - W4 & 257$^{+16}_{-16}$ / 340 (75.6$^{+4.7}_{-4.7}$\%) & 40$^{+6}_{-6}$ / 72 (55.6$^{+8.3}_{-8.3}$\%) & 15$^{+4}_{-4}$ / 332 (4.5$^{+1.2}_{-1.2}$\%) \\
Radio (FIRST) & 27$^{+5}_{-5}$ / 346 (7.8$^{+1.4}_{-1.4}$\%) & 8$^{+4}_{-3}$ / 73 (11.0$^{+5.5}_{-4.1}$\%) & 3$^{+3}_{-2}$ / 333 (0.9$^{+0.9}_{-0.6}$\%) \\
X-ray \& Radio & 3$^{+3}_{-2}$ / 346 (0.9$^{+0.9}_{-0.6}$\%) & 1$^{+2}_{-1}$ / 73 (1.4$^{+2.8}_{-1.4}$\%) & 0 / 333 (0.0\%)\\
\hline
\end{tabular}
\end{table*}

\subsection{Multiwavelength Data in X-ray, MIR, and Radio} \label{sec:surveys}



We match our samples to public catalogs to identify detected galaxies and summarize the detection rates in Table \ref{tab:det_rate}. In the X-ray, we use the eROSITA/eFEDS MAIN point source counterparts catalog \citep{Brunner_2022,Salvato_2022} to search for detections in our samples. We match all our galaxies using a matching radius of 1\arcsec to the optical positions of the X-ray counterparts from the catalog (\texttt{SPECZ\_RA} and \texttt{SPECZ\_DEC}). We regard galaxies that have matches in the eROSITA/eFEDS main catalog as detected in X-ray. We obtain the X-ray properties for detected sources from the eROSITA/eFEDS X-ray spectral catalog \citep{Liu_2022} which contains X-ray spectral analysis for all the eFEDS sources. Given the low detection rates in X-ray, we perform imaging stacking (Section \ref{sec:xray stacking}) using eFEDS images published as part of the eROSITA Early Data Release (EDR)\footnote{\url{https://erosita.mpe.mpg.de/edr/eROSITAObservations/Catalogues/liuT/eFEDS_c001_images/}}. 

In the MIR, we utilize the unWISE catalog \citep{Lang_2016} which provides forced photometry measurements at SDSS galaxy locations. We use a matching radius of 1\arcsec and define detections as SNR $\equiv$ $\texttt{w?\_nanomaggies} \times \sqrt{\texttt{w?\_nanomaggies\_ivar}} >$ 3 using corresponding columns from the catalog (``\texttt{w?}" refers to the WISE bands W1, W2, W3, and W4 at 3.4$\mu$m, 4.6$\mu$m, 12$\mu$m, 22$\mu$m). A few galaxies end up near masked bright stars in the unWISE images and therefore have no match from the catalog. We obtain measurements for 72 PSBs, 340 SFGs, and 332 QGs from the unWISE catalog and note these numbers in Table \ref{tab:det_rate}.

In the radio, we cross-match all our galaxies to the 1.4 GHz FIRST catalog (14Dec17 Version, \citealt{Becker_1995,Helfand_2015}) with a matching radius of 2\arcsec. We regard galaxies that have matches in the FIRST catalog as detected in radio emission, and use their integrated flux densities from the catalog for later analysis. We perform image stacking as described in Section \ref{sec:radio stacking} using FIRST image cutouts obtained via \texttt{astroquery}. The Karl G. Jansky Very Large Array Sky Survey (VLASS, \citealt{lacy2020}) at 3 GHz is another all-sky radio survey. While VLASS reaches greater depth than FIRST, our galaxies have lower detection rates in the VLASS Quick Look epoch 2 catalog\footnote{\url{https://cirada.ca/vlasscatalogueql0}} \citep{Gordon_2021}, likely due to intrinsically fainter emission at 3 GHz (L$_{\rm 3 GHz}$/L$_{\rm 1.4 GHz}$ $\sim$ 0.6 assuming a typical spectral index of $-$0.7 \citep[e.g,][]{Condon_1992}) and the quality of Quick Look images. VLASS Single Epoch continuum images which are better cleaned and flux-calibrated than Quick Look images are not yet available for all of our galaxies. We therefore use FIRST data for our radio stacking analysis.


\begin{figure*}
\centering
\includegraphics[width=\textwidth]{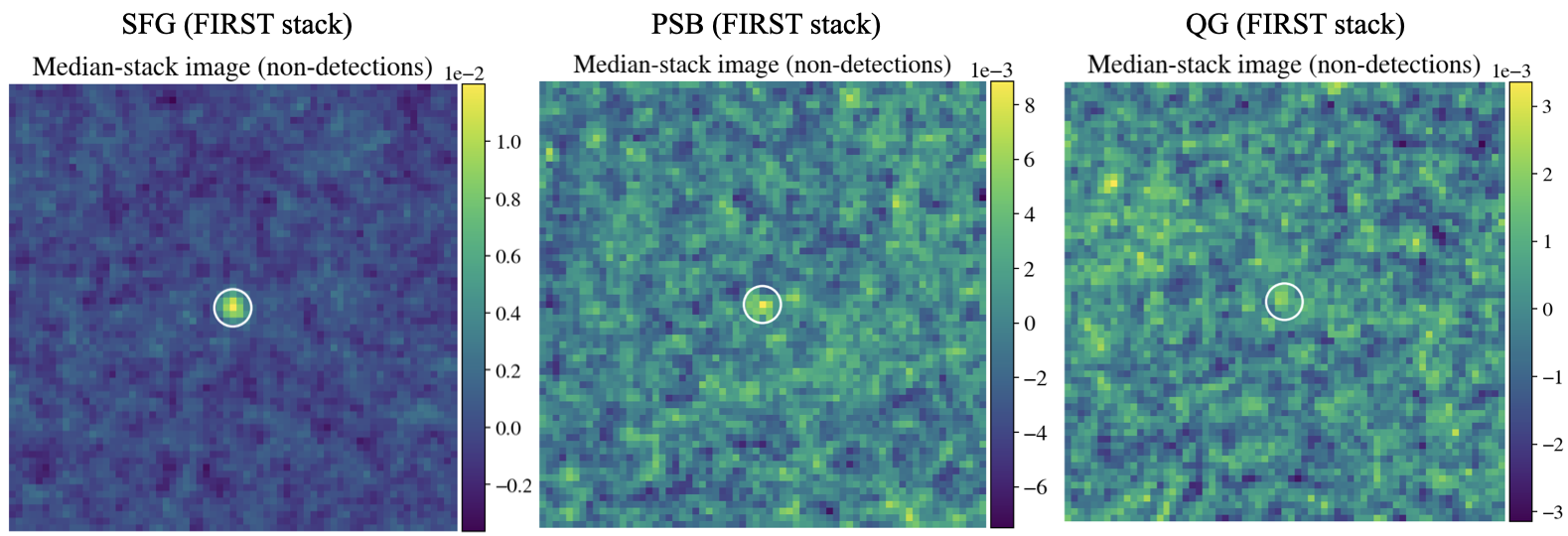}
\caption{Median stacked radio images (2\arcmin $\times$ 2\arcmin, 1.8\arcsec/pixel) in mJy for undetected SFGs, PSBs, and QGs. The white circle represents the 10\arcsec\ diameter aperture used to measure the stacked signal (Table \ref{tab:radio_stack}). All three samples show SNR $>$ 2 signals in the stacks.
\label{fig:radio-stack}}
\end{figure*}

\begin{figure}
\centering
\includegraphics[width=\columnwidth]{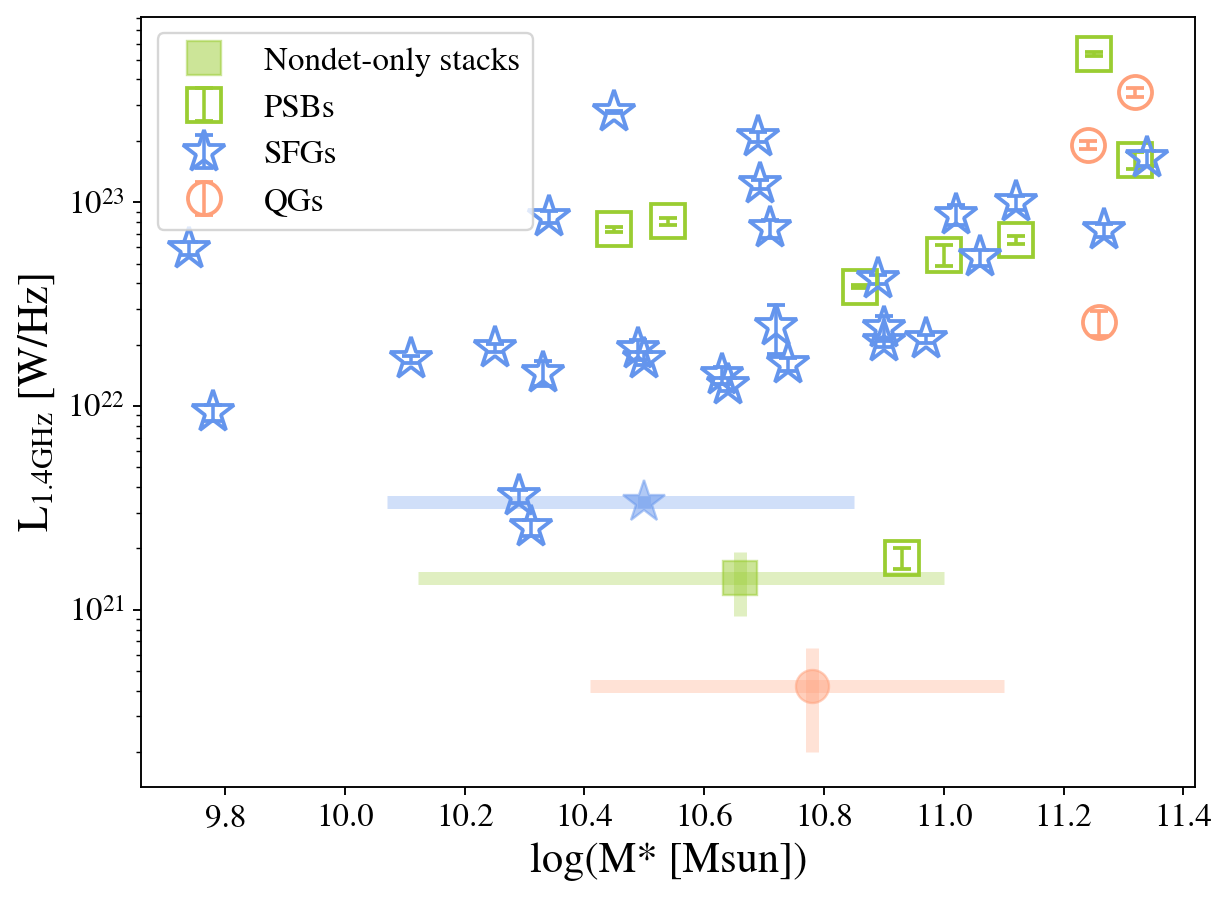}
\caption{Open points are galaxies detected in the FIRST catalog and filled points are measurements from stacking only undetected galaxies. The horizontal bar on the stacked points shows the 16th and 84th percentile range of the M$_*$ and the vertical bar shows the luminosity error from the bootstrapping method. Luminosity errors for individually detected galaxies (from the RMS in the catalog) are plotted though the error bars are smaller than the marker size.
\label{fig:l_rad}}
\end{figure}

\section{Analysis} \label{sec:analysis}

The majority of our galaxies are not individually detected in radio and X-ray (Table \ref{tab:det_rate}). PSBs show approximately twice the detection rates of SFGs and QGs in radio and X-ray, though the detection rates for all three samples may still be consistent given the large uncertainties. Among the 6 BPT-AGN PSBs, 1 is detected in X-ray and 2 in radio, which are within the counting uncertainties as listed in Table \ref{tab:det_rate}. Thus a small number of optically identified AGN within the PSB sample does not significantly bias the overall radio and X-ray detection rates for PSBs. In the MIR, all galaxies are detected in the W1 and W2 bands, with detection rates decreasing from W2 to W4 in all samples. This decrease could result from both observational depth variations in different bands and changes in intrinsic SED shape (more in Section \ref{sec:average mir}).


The low radio and X-ray detection rates observed for PSBs in our study are consistent with previous findings, which report that only $\sim$ 2\% of low-redshift PSBs are detected in FIRST \citep[e.g.,][]{Shin_2011,Mullick_2023} and that most exhibit fewer than ten photons in \textit{Chandra} X-ray observations (corresponding to $\lesssim$ 10$^{42}$ erg/s, e.g., \citealt{Brown_2009,Lanz_2022}). We perform image stacking for the non-detections to recover a signal representative of each galaxy sample, as detailed in the following sections. By stacking individually undetected sources, we minimize bias from a small number of relatively luminous, detected galaxies. The narrow redshift range of our sample ($z < 0.2$) ensures that redshift-dependent effects have a negligible influence on the stacking results. While the galaxies undetected in radio are not entirely the same as those undetected in X-ray, the overall small detection rates at both wavelengths ensure that the radio and X-ray stacks for each sample contain $\gtrsim$ 90\% identical galaxies and are thus representative of their respective galaxy populations. 



\subsection{Radio Stacking} \label{sec:radio stacking}



We create a stacked image for the undetected galaxies in each SFG, PSB, and QG sample. Within each sample, we extract 2\arcmin $\times$ 2\arcmin\ image cutouts centered on the galaxy of interest and combine the cutouts following the median stacking method in \citet{White_2007}. We randomly rotate the cutouts before stacking to mitigate the effects of stripy artifacts due to image deconvolution. We perform aperture photometry with circular apertures twice the beam FWHM (10\arcsec\ diameter) on the stacked images to obtain fluxes in the stack. 

We assess the robustness of the stacked signal by comparing the peak pixel flux within the aperture to the root-mean-square (rms) noise measured from pixels of the stacked image outside the source aperture\footnote{Following the SNR calculation convention in the FIRST survey (\url{https://sundog.stsci.edu/first/catalogs/readme.html}) but without the 0.25 mJy correction in the original equation for our stacked images.}. All three samples show SNR $>$ 2 signals in the stacks\footnote{SFG stack SNR $\gtrsim$ 13; PSB stack SNR $\gtrsim$ 4; QG stack SNR $\gtrsim$ 2.}. To account for both photometric and systematic statistical errors associated with stacking, we perform a bootstrapping analysis to estimate the stacked flux uncertainty. For each stacked sample, we create 500 mock samples of the same size by randomly selecting from the stacked sample with replacement. We follow the same procedures to obtain the stacked fluxes in the mock samples and take the standard deviation of the resulting flux distributions as the error on our stacked flux measurements. Our stacked radio images and flux measurements are presented in Figure \ref{fig:radio-stack} and Table \ref{tab:radio_stack}.

We convert the stacked fluxes at 1.4 GHz to luminosity using the median redshifts within the stacks and plot them with individually detected galaxies in Figure \ref{fig:l_rad}. The individually detected galaxies follow the general trend of increasing L$_{\rm 1.4GHz}$ with increasing M$_*$ and the three galaxy populations show a significant overlap in L$_{\rm 1.4GHz}$. PSBs have an intermediate stacked radio luminosity between SFGs and QGs. There are only a few galaxies in each sample that fall in the traditional radio-loud regime of L$_{\rm 1.4GHz} >$10$^{23}$ W/Hz \citep{Best_2005_sample_of_radio_loud_agn,Best_2005_host_galaxies_radio-loud_agn}. We look further into the possibility of radio emission from AGN in Section \ref{sec:agn in radio}.

\begin{table*}
\centering
\caption{Flux and luminosity measurements from stacked FIRST cutouts of undetected galaxies. The luminosity is converted from the flux using the median redshift in the stack. The median SFR is the median of the values for galaxies that are included in the stack.\label{tab:radio_stack}}
\begin{tabular}{lcccc}
\hline\hline
Sample & Stacked flux [$\mu$Jy] & Stacked luminosity [W/Hz] & Median log$_{10}$(SFR [M$_{\odot}$/yr]) in stack \\
\hline
SFG & \( 157 \pm 11 \, (0.88) \) & (34 $\pm$ 2.4) $\times$ 10$^{20}$ & 0.5 \\
PSB & \( 69 \pm 24 \, (1.9) \) & (14 $\pm$ 5.0) $\times$ 10$^{20}$ & 0.03 \\
QG  & \( 19 \pm 10 \, (0.84) \) & (4.2 $\pm$ 2.2) $\times$ 10$^{20}$ & $-$3.7 \\
\hline
\end{tabular}
\tablecomments{Flux and luminosity uncertainties in this table are from bootstrapping (Section \ref{sec:radio stacking}). Values inside the parentheses for the stacked fluxes are the stacked image pixel rms.}
\end{table*}

\begin{deluxetable*}{cccccccccc}
\renewcommand{\arraystretch}{1.3} 
\tablecaption{Median luminosity [10$^9$L$_\odot$] and flux density [mJy] for our samples of SFGs, PSBs, and QGs in the four WISE bands. Median SFR in the corresponding samples are also included. Numbers in the subscripts and superscripts are the 16th and 84th percentiles. \label{tab:mir-stack}}
\tablehead{
    \colhead{Sample} & 
    \colhead{$\nu$L$_{\rm W1}$} & \colhead{$\nu$L$_{\rm W2}$} & \colhead{$\nu$L$_{\rm W3}$} & \colhead{$\nu$L$_{\rm W4}$} & \colhead{F$_{\rm W1}$} & \colhead{F$_{\rm W2}$} & \colhead{F$_{\rm W3}$} & \colhead{F$_{\rm W4}$} & \colhead{\makecell{Median\\log$_{10}$(SFR [M$_{\odot}$/yr])}}
}
\startdata
SFG (all) & 5.9$^{13.1}_{2.3}$ & 3.0$^{7.5}_{1.1}$ & 6.1$^{15.8}_{2.1}$ & 4.5$^{11.9}_{1.0}$ & 0.98$^{2.0}_{0.59}$ & 0.71$^{1.4}_{0.41}$ & 4.0$^{8.7}_{1.9}$ & 5.2$^{11.9}_{2.0}$ & 0.5\\
PSB (all) & 5.9$^{15.1}_{1.8}$ & 3.0$^{8.4}_{0.9}$ & 2.7$^{10.5}_{0.5}$ & 2.4$^{8.5}_{0.0}$ & 0.83$^{2.3}_{0.53}$ & 0.56$^{1.9}_{0.32}$ & 1.7$^{7.3}_{0.5}$ & 3.7$^{13.0}_{0.0}$ & $-$0.04 \\
QG (all)  & 5.3$^{13.0}_{2.2}$ & 2.3$^{6.4}_{0.9}$ & 0.3$^{1.3}_{0.0}$ & 0.03$^{1.5}_{0.0}$ & 0.78$^{1.5}_{0.57}$ & 0.49$^{0.86}_{0.34}$ & 0.20$^{0.75}_{0.0}$ & 0.06$^{1.5}_{0.0}$ & $-$3.7 \\
SFG (SNR $>$ 3) & \nodata\tablenotemark{a} & \nodata & 6.1$^{15.8}_{2.1}$ & 6.0$^{15.7}_{2.4}$ & \nodata & \nodata & 4.0$^{8.7}_{1.9}$ & 6.6$^{14.5}_{3.8}$ & 0.5; 0.6\tablenotemark{b} \\
PSB (SNR $>$ 3) & \nodata & \nodata & 3.2$^{11.5}_{1.1}$ & 5.1$^{22.3}_{1.7}$ & \nodata & \nodata & 1.9$^{7.9}_{0.8}$ & 7.2$^{24.8}_{3.8}$ & 0.1; 0.4 \\
QG (SNR $>$ 3)  &\nodata & \nodata & 1.3$^{3.5}_{0.4}$ & 2.3$^{10.3}_{0.7}$ & \nodata & \nodata & 0.86$^{1.8}_{0.54}$ & 4.1$^{5.8}_{2.9}$ & $-$3.6; $-$3.7 \\
\enddata
\tablenotetext{a}{All galaxies have SNR $>$ 3 in the W1 and W2 bands and corresponding values are not repeated in the table.}
\tablenotetext{b}{The two numbers correspond to median SFRs of galaxies with SNR $>$ 3 in the W3 and W4 bands, respectively.}
\end{deluxetable*}

\subsection{Composite MIR SED} \label{sec:average mir}

\begin{figure}
\centering
\includegraphics[width=\columnwidth]{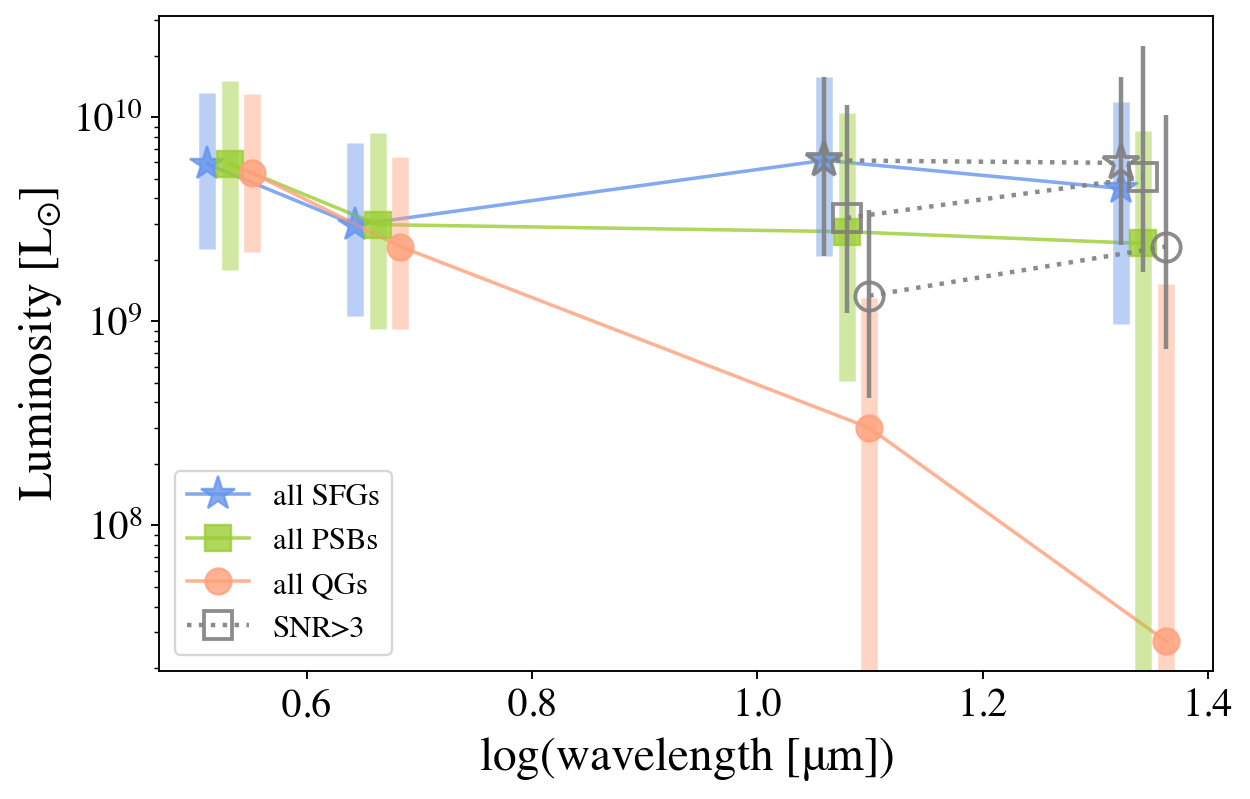}
\caption{The composite MIR SED for SFGs, PSBs, and QGs in the 4 WISE bands: W1 (3.4$\mu$m), W2 (4.6$\mu$m), W3 (12$\mu$m), W4 (22$\mu$m). Each symbol represents the median luminosity of the corresponding sample converted from nanomaggies, with the vertical bars showing the 16th and 84th percentiles. All samples are detected with SNR $>$ 3 in W1 and W2 bands, and we overplot in gray open points the medians and percentiles of the subsamples with SNR $>$ 3 in W3 and W4 bands.
\label{fig:mir_sed}}
\end{figure}

Since a great number of galaxies are well-detected in the MIR, we use the forced photometry from the unWISE catalog to explore the average MIR properties across our samples instead of image stacking. We construct the composite MIR SED for SFGs, PSBs, and QGs using the four WISE bands at 3.4$\mu$m (W1), 4.6$\mu$m (W2), 12$\mu$m (W3), 22$\mu$m (W4) by taking the median luminosity of the sample converted from nanomaggies. We note that although a few galaxies individually detected in radio or X-ray show AGN signatures at those wavelengths (Section \ref{sec:agn in radio}, \ref{sec: agn in xray}), they are unlikely to bias the overall shapes of the MIR composite SEDs, as the latter take into account the full samples containing hundreds of galaxies. The composite MIR SED is plotted in Figure \ref{fig:mir_sed} and the corresponding luminosity values are recorded in Table \ref{tab:mir-stack}.

All three samples show similar slopes (colors) between W1 and W2. For the composite SEDs including all galaxies, PSBs show a relatively flat luminosity gradient from W2 to W4, SFGs are slightly brighter in W3 than W4, while QGs display a dramatic drop. The overall low detection rates of QGs in W3 (26.8\%) and W4 (4.5\%), along with their low luminosities in these bands, are consistent with previous findings that early-type galaxies exhibit little emission at these wavelengths \citep[e.g.,][]{Jarrett_2019}. PSBs and SFGs show substantially higher detection rates in W3 and W4 ($\gtrsim50$\%; see Table \ref{tab:det_rate}). If we only look at galaxies with well-detected W3 and W4 (gray open points on Figure \ref{fig:mir_sed}), SFGs show a flat gradient between W3 and W4 while both PSBs and QGs have larger W4 luminosity than their W3 luminosity. Composite MIR SEDs of similar shapes have been reported in \citet{Alatalo_2017} within samples of nearby PSBs, early-, and late-type galaxies.

W1 and W2 trace emission from low-mass, older stars and have been used to derive stellar masses for galaxies \citep[e.g.,][]{Jarrett_2013,Wen_2013}. Our samples are matched in M$_*$ so it is not unexpected that they have comparable luminosities in W1 and W2. The primary feature in the W3 band is the polycyclic aromatic hydrocarbon (PAH) emission at 11.3 $\mu$m, which traces SF \citep[e.g.,][]{Shipley_2016}. Warm continuum emission from dust heated by AGN or young stars could also contribute to the observed luminosity in W3. The W2--W3 color is a probe for the present level of star formation \citep[e.g.,][]{Mateos_2012,Coziol_2014} and the ratio between stellar and hot dust luminosities \citep{Alatalo_2017}. There are no PAH features in the W4 band and it is dominated by the continuum emission from hot dust grains, which could come from AGN activity and dusty starbursts \citep[e.g.,][]{Jarrett_2013}. Since our galaxies are unresolved in WISE, the MIR emission consists of both nuclear activity and host galaxy SF. The decrease in W3 luminosity and in the W2--W3 color from SFGs to QGs could reflect a decrease in SF and the corresponding SF dominance in the total luminosity we observe. A similar decreasing trend in W3 luminosity with starburst age was also observed in \citet{Rowlands_2015}. Thermally pulsating asymptotic giant branch (TP-AGB) stars and their stellar winds could enrich the ISM with carbonaceous material for PAH emission, and associated shocks might preferentially destroy smaller PAH molecules that emit at shorter wavelengths (e.g., 6--9 $\mu$m), leading to relatively stronger PAH emission at 11.3 $\mu$m \citep{Vega_2010}. Since PSBs harbor abundant intermediate age stars, their elevated W3 luminosities relative to QGs may imply partial contributions from AGB stars (also see \citealt{Alatalo_2017}). However, any such AGB star contributions are unlikely to be substantial given that PSBs do not show paticularly high W3 luminosities (still lower than those of SFGs).
The rise in W4 luminosity compared to W3 in detected PSBs and QGs could be due to warmer dust heated by AGN or compact starbursts. The starburst scenario is unlikely for our QGs since they contain old stellar populations and have long been quenched. The past starbursts in some of our PSBs at the very early quenching stage could potentially affect their MIR SED shape. We look further into the possibility of AGN utilizing MIR colors in Section \ref{sec: agn in mir}.




\subsection{X-ray Stacking} \label{sec:xray stacking}






We generate count-rate stacked images using the count and exposure maps in the 0.2--2.3 keV energy range, where eROSITA is most sensitive. We follow the stacking procedure in \citet{Toba_2022}. Prior to stacking, we mask sources previously detected in the eROSITA/eFEDS main catalog \citep{Brunner_2022,Salvato_2022} using a circular mask with a radius equal to the cataloged source extent when the extent exceeds 30\arcsec, and 30\arcsec\ otherwise. The X-ray images have a resolution of 4\arcsec/pixel and we create 640\arcsec $\times$ 640\arcsec\, cutouts centered on our galaxies. For each sample (SFGs, PSBs, and QGs), we construct a stacked image by taking the mean of the count-rate cutouts, which are obtained by dividing the count map cutouts by the corresponding exposure time map cutouts. We require the mean exposure time in the central 1\arcmin $\times$ 1\arcmin\ region to be greater than 180s for a cutout to be included in the stack. This will remove galaxies that fall near the edge of the eFEDS field and improve the quality of the final stacked images. The fraction of galaxies removed by this criterion is $\lesssim$ 7\%.

\begin{figure*}
\centering
\includegraphics[width=\textwidth]{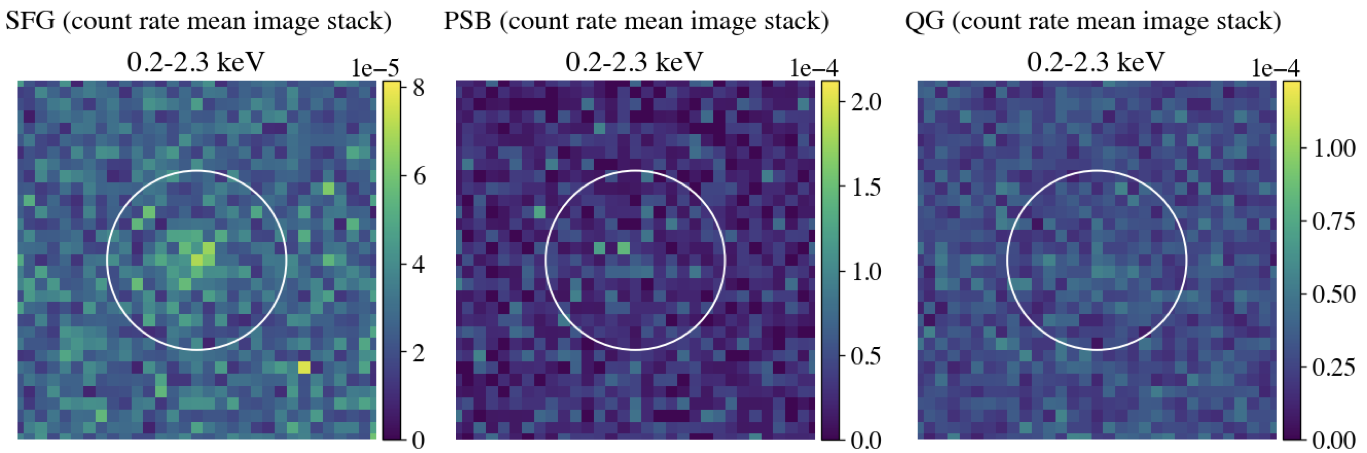}
\caption{Mean stacked count rate X-ray images (2\arcmin $\times$ 2\arcmin, 4\arcsec/pixel) from eROSITA for undetected SFGs, PSBs, and QGs. The white circle represents the 30\arcsec\, radius aperture used to measure the stacked signal (Table \ref{tab:x-stack}). We detect signals with SNR $>2$ in the SFG and PSB stacks, while no significant signal is detected in the QG stack.
\label{fig:xstack}}
\end{figure*}

We measure the signal at the center of the stacked count-rate images with a 30\arcsec\ radius aperture. To correct for the background, we randomly measure the signal from 10 off-center apertures of the same size and subtract the median count rate from the source measurement. The uncertainty in the mean-stacked count rates is propagated from the Gehrels uncertainty\footnote{$\delta C=1+\sqrt{C+0.75}$, where C is the total counts within the source aperture and $\delta C$ is the corresponding Gehrels uncertainty \citep{Gehrels_1986}.} \citep{Gehrels_1986} in the individual count images assuming no errors on the exposure times and background. The background-subtracted count rates and corresponding errors are converted to observed fluxes in the 0.2--2.3 keV band using WebPIMMS\footnote{\url{https://heasarc.gsfc.nasa.gov/cgi-bin/Tools/w3pimms/w3pimms.pl}} assuming a power law model with photon index $\Gamma = 1.7$ (following \citealt{Toba_2022}, which is a typical value for AGN, see, e.g., \citealt{Marchesi_2016}) and Galactic absorption N$_{\rm H,Galactic}$ = 3 $\times$ 10$^{20}$ cm$^{-2}$ (as for the eFEDS catalog in \citealt{Brunner_2022}). For individually detected galaxies in the \citet{Brunner_2022} catalog, we convert their observed fluxes and flux errors in 0.5--2\,keV from the single power-law fitting to 0.2--2.3\,keV using WebPIMMS with their posterior median $\Gamma$ from the \citet{Liu_2022} catalog. Our stacked count rate images and flux measurements are presented in Figure \ref{fig:xstack} and Table \ref{tab:x-stack}. We detect signals with SNR $>2$ in the stacks for SFGs and PSBs. We do not detect a significant signal in the QG stack and use the 2$\sigma$ upper limit for the QG stack in later analysis.

\begin{figure*}
\centering
\includegraphics[width=\textwidth]{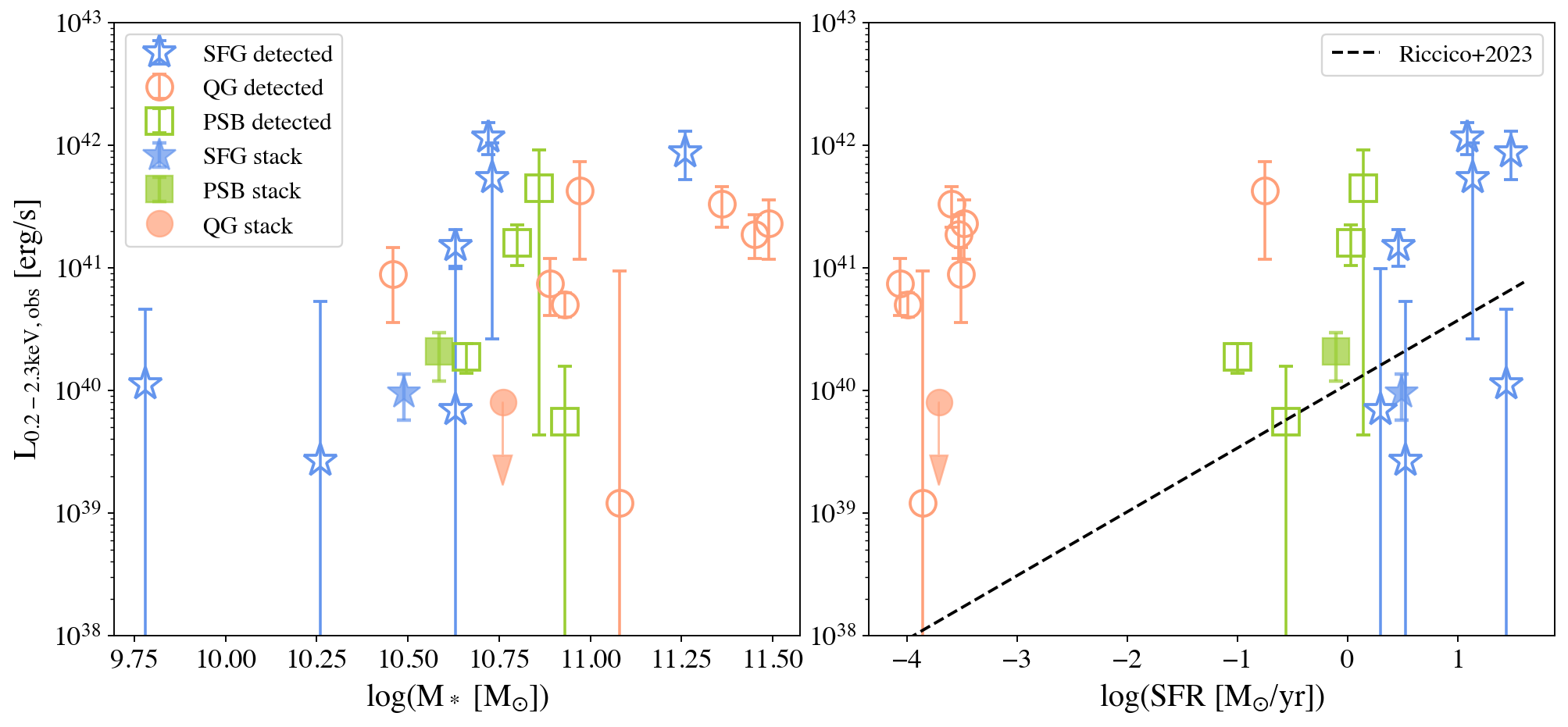}
\caption{\textbf{Left}: Observed X-ray luminosity vs stellar mass. \textbf{Right}: Observed X-ray luminosity vs SFR. Open points represent individually detected sources in the eFEDS catalog and filled points represent measurements from the stacked images (2$\sigma$ upper limit for the QG stack). All three samples show a variety of X-ray luminosity, with no single population displaying systematically larger X-ray luminosity than the others. The relation between L$_{\rm X}$ vs SFR from the literature is plotted for reference, though our galaxies don't exhibit a clear correlation between these two quantities.
\label{fig:lx_ms_sfr}}
\end{figure*}

We convert the stacked fluxes in 0.2--2.3\,keV to luminosity using the median redshifts within the stacks and plot L$_{\mathrm{0.2-2.3keV,obs}}$ against M$_*$ and SFR in Figure \ref{fig:lx_ms_sfr}. Individually detected PSBs are not systematically brighter in the X-ray than SFGs and QGs. The PSB stack shows marginally larger X-ray luminosities than the SFG and QG stacks, though they are consistent within the uncertainties. Similar to the radio observations, the majority of our galaxies (and stacks) are faint in X-ray ($<$10$^{43}$ erg/s). More than half of the individually detected galaxies, and the PSB and QG stacks lie above the L$_{\rm X}$-SFR relation derived from the same eFEDS observation used in this work \citep{Riccio_2023}, suggesting excess X-ray emission from AGN.  We look further into the possibility of AGN in X-ray in Section \ref{sec: agn in xray}.


\begin{table*}
\centering
\caption{Count rate measurements from stacked eROSITA cutouts of undetected galaxies, and the corresponding flux and luminosity in 0.2--2.3 keV. The luminosity is converted from the flux using the median redshift in the stack. The median SFR is the median of the values for galaxies that are included in the stack. \label{tab:x-stack}}
\begin{tabular}{lcccc}
\hline\hline
Sample & Count rate [cts s$^{-1}$] & Flux [erg s\(^{-1}\) cm\(^{-2}\)] & Luminosity [erg s\(^{-1}\)] & Median log$_{10}$(SFR [M$_{\odot}$/yr]) in stack \\
\hline
SFG & (4.7 $\pm$ 1.9) $\times$ 10$^{-4}$ & (4.5 $\pm$ 1.8) $\times$ 10$^{-16}$ & (9.6 $\pm$ 3.9) $\times$ 10$^{39}$ & 0.5 \\
PSB & (11.0 $\pm$ 4.7) $\times$ 10$^{-4}$ & (10.3 $\pm$ 4.4) $\times$ 10$^{-16}$ & (20.8 $\pm$ 8.8) $\times$ 10$^{39}$ & $-$0.1 \\
QG  & $<$ 4.1 $\times$ 10$^{-4}$ & $<$ 3.8 $\times$ 10$^{-16}$ & $<$ 8.0 $\times$ 10$^{39}$ & $-$3.7 \\
\hline
\end{tabular}
\tablecomments{Flux conversion assumes Galactic absorption N$_{\rm H,Galactic}$ = 3 $\times$ 10$^{20}$ cm$^{-2}$ (as for the eFEDS catalog in \citealt{Brunner_2022}). QGs are not significantly detected in the stack, and the table numbers represent the 2$\sigma$ upper limits.}
\end{table*}

\section{Multiwavelength Perspectives of AGN} \label{sec: multwavelength agn}

In this section, we investigate the evidence for AGN across various wavelengths for each galaxy sample and discuss the implications of these findings in the broader context of SF quenching in Section \ref{sec:discussion}. To evaluate the contribution from AGN to a galaxy's light, we must estimate the emission due to SF utilizing SFR. When available, we adopt the SFR values from the \citet{Chang_2015} catalog, which are based on the SED fitting of combined SDSS and WISE photometry. \citet{Chang_2015} consistently modeled both the attenuated stellar SED and the dust emission at 12 and 22 $\mu$m. Their resulting SFRs taking into account PAH emission and thermal dust radiation and are thus robust to obscured SF as well. Optical-MIR-based SFRs are sensitive to SF over relatively longer timescales ($\lesssim 100$ Myr) and could be affected by dust primarily heated by old stars \citep[e.g.,][]{daCunha_2008,Kennicutt_2012}. Therefore, these SFRs might be overestimated for PSBs due to their recent starburst and relatively low current star formation level \citep[as seen in, e.g.,][]{Salim_2016,Wild_2025}.

A small fraction ($\sim$ 1\%) of our galaxies don't have SFRs from \citet{Chang_2015} and we obtain their SFRs from the MPA/JHU catalog. \citet{Chang_2015} observed a median offset of 0.22 dex between their SFR values and those from the MPA-JHU catalog due to different SF tracers used. However, this difference should not affect our analysis as we consistently adopt the Chang SFRs for $\sim$ 99\% of our galaxies and use the median SFRs for the stacks. We include MPA-JHU SFRs for the small number of galaxies without Chang SFRs for completeness. SFRs from the MPA-JHU catalog are based on emission lines for their classified SFGs and photometry for their classified AGN and galaxies with weak emission lines \citep{Brinchmann_2004}. We use their estimated total SFR after aperture corrections.



\subsection{AGN vs. SF in Radio} \label{sec:agn in radio}

\begin{figure}
\centering
\includegraphics[width=\columnwidth]{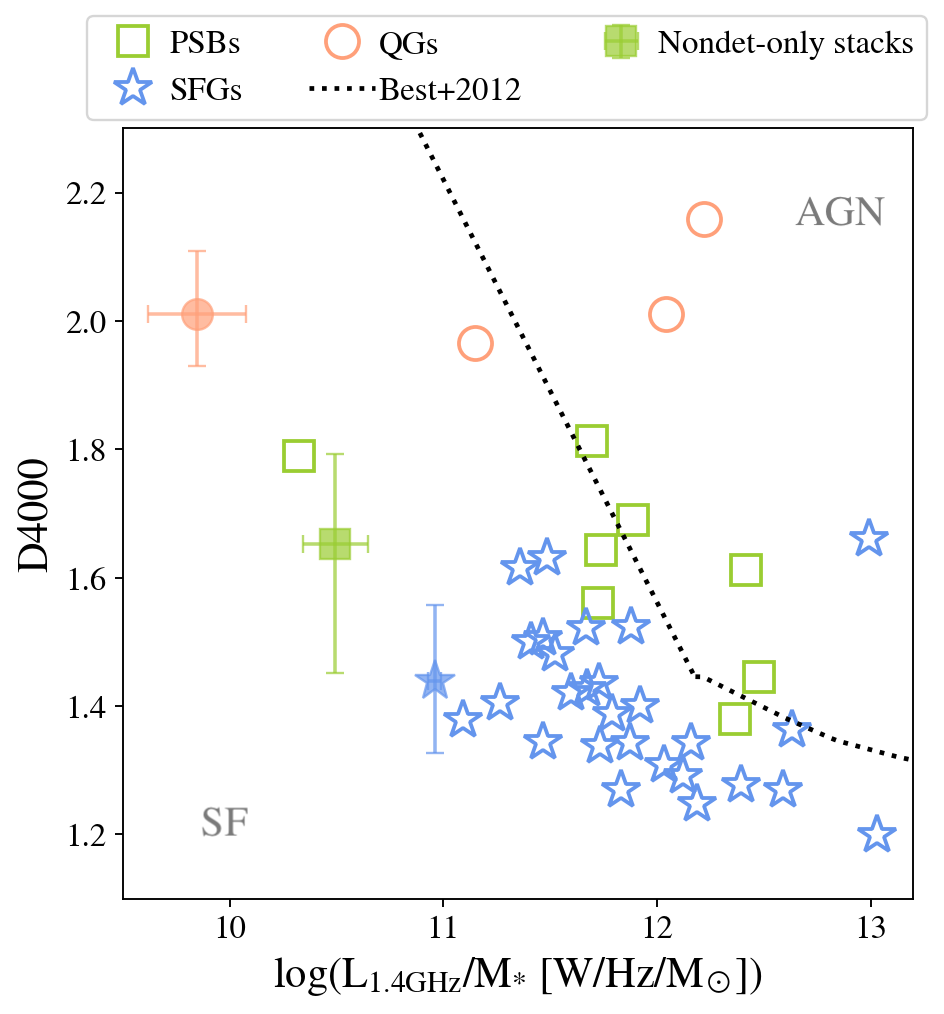}
\caption{Dn4000 versus radio luminosity per stellar mass with the SF/AGN demarcation line from \citet{Best2012}. Points below the dotted line are supposed to have radio emission powered by SF. The vertical error bars on the stacked points (filled) show the 16th and 84th percentile range of the Dn4000 values within the stack and the horizontal error bars show propagated errors from the stacked luminosity error with the bootstrapping method (M$_*$ are the medians in the stacks). The stacked measurements fall in the region where SF dominates radio luminosity. Individually detected PSBs lie around the demarcation line and between individually detected SFGs and QGs. 
\label{fig:best_radio}}
\end{figure}

\citet{Best2012} and \citet{Best_2005_sample_of_radio_loud_agn} plotted Dn4000 versus radio luminosity per stellar mass to separate radio-loud AGN from galaxies where the radio emission is powered by SF. We plot the same quantities for our galaxies in Figure \ref{fig:best_radio}. Despite only a few detections, QGs lie primarily on the AGN side while SFGs lie primarily on the SF side. Individually detected PSBs lie around the demarcation line and between individually detected SFGs and QGs. This observation reflects the transitioning nature of our PSBs and that both SF and AGN may contribute to their radio emission. The stacked measurements, on the other hand, fall in the SF region for all three samples.

\begin{figure}
\centering
\includegraphics[width=\columnwidth]{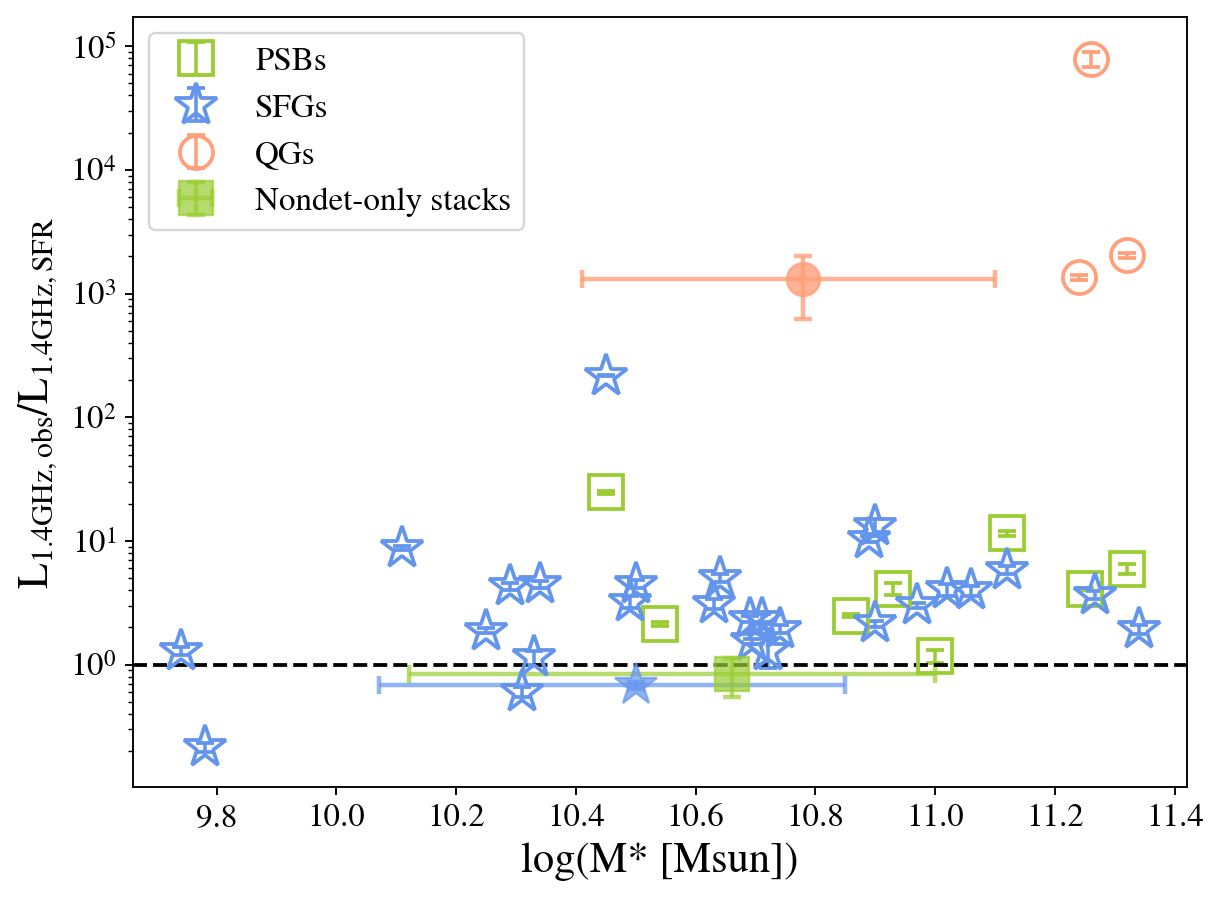}
\caption{L$_{\mathrm{1.4GHz,obs}}$/L$_{\mathrm{1.4GHz,SFR}}$ versus stellar mass. Open points are galaxies detected in the FIRST catalog and filled points are measurements from stacking only undetected galaxies. The horizontal error bars on the stacked points shows the 16th and 84th percentile range of the M$_*$ and the vertical error bars show propagated errors from the stacked luminosity error with the bootstrapping method. The vertical errors for the individually detected galaxies are propagated from uncertainties in L$_{\mathrm{1.4GHz,obs}}$ from the FIRST catalog. See Section \ref{sec:agn in radio} for more discussions on uncertainties. PSBs and the majority of SFGs do not show significantly enhanced L$_{\mathrm{1.4GHz,obs}}$/L$_{\mathrm{1.4GHz,SFR}}$.
\label{fig:lrad_ms}}
\end{figure}

To further relate the observed luminosity with star formation, we estimate the radio luminosity expected from star formation using the SFR for each galaxy (mostly from \citealt{Chang_2015}, median SFR for stacks) by inverting the relation from \citet{Murphy_2011}:
\begin{align}
\left( \frac{L_{\rm 1.4\,GHz, SFR}}{\mathrm{erg}\, \mathrm{s}^{-1}\, \mathrm{Hz}^{-1}} \right) = \left( \frac{1}{6.35 \times 10^{-29}} \right) \left( \frac{\mathrm{SFR}}{M_{\odot}\, \mathrm{yr}^{-1}} \right).
\end{align}
We plot the ratio L$_{\mathrm{1.4GHz,obs}}$/L$_{\mathrm{1.4GHz,SFR}}$ with M$_*$ in Figure \ref{fig:lrad_ms}. Despite some scatter, radio-detected SFGs and PSBs show moderately enhanced L$_{\mathrm{1.4GHz,obs}}$/L$_{\mathrm{1.4GHz,SFR}}$ ($\sim$ 5) indicating some possible radio emission from AGN, while the SFG and PSB stacks are consistent with the scenario where SF is responsible for the radio emission. Accounting for the uncertainty in the SFR$_{\mathrm{1.4GHz}}$ from the \citet{Murphy_2011} relation, which is $\sim$0.6 dex compared to their standard SFR$_{\mathrm{33GHz}}$, the observed moderate enhancement in L$_{\mathrm{1.4GHz,obs}}$/L$_{\mathrm{1.4GHz,SFR}}$ is likely not significant. In addition, radio emission at this low frequency is sensitive to SF over longer timescales up to several hundred Myr \citep[e.g.,][]{Arango_Toro_2023,Cook_2024}. Thus the expected L$_{\mathrm{1.4GHz}}$ from SF in PSBs might be underestimated when derived from optical-MIR-based SFRs, given their bursty SF histories. In this context, the lack of elevated radio AGN activity in PSBs compared to SFGs -- and the possibility of even lower L$_{\mathrm{1.4GHz,obs}}$/L$_{\mathrm{1.4GHz,SFR}}$ ratios -- would be fully consistent with a scenario in which all radio emission originates from SF. A similar absence of radio AGN signatures in $z>0.5$ PSBs has also been reported in Patil et al. (in prep). QGs (individually detected and stacked) display significantly higher L$_{\mathrm{1.4GHz,obs}}$/L$_{\mathrm{1.4GHz,SFR}}$ on the order of $\gtrsim$ 1000. Given the minimal SFR of QGs, all of their radio emission may come from AGN. The comparison of L$_{\mathrm{1.4GHz,obs}}$ and L$_{\mathrm{1.4GHz,SFR}}$ implies that QGs host relatively stronger radio AGN than SFGs and PSBs. Radio undetected SFGs and PSBs might not contain AGN at all. These observations are consistent with the implication from Figure \ref{fig:best_radio} that none of the stacks show AGN-dominated radio emission. We note the statistical constraint that QGs have the smallest radio detection rate and faintest stacked radio luminosity ($\lesssim$ 10$^{21}$ W/Hz). Thus the QG stack also may not contain AGN or only very weak ones. This could explain why the QG stack appears on the SF side of Figure \ref{fig:best_radio}, given that the demarcation in the figure is designed for radio-loud AGN.


\subsection{AGN vs. SF in MIR} \label{sec: agn in mir}

\begin{figure*}
\centering
\includegraphics[width=\columnwidth]{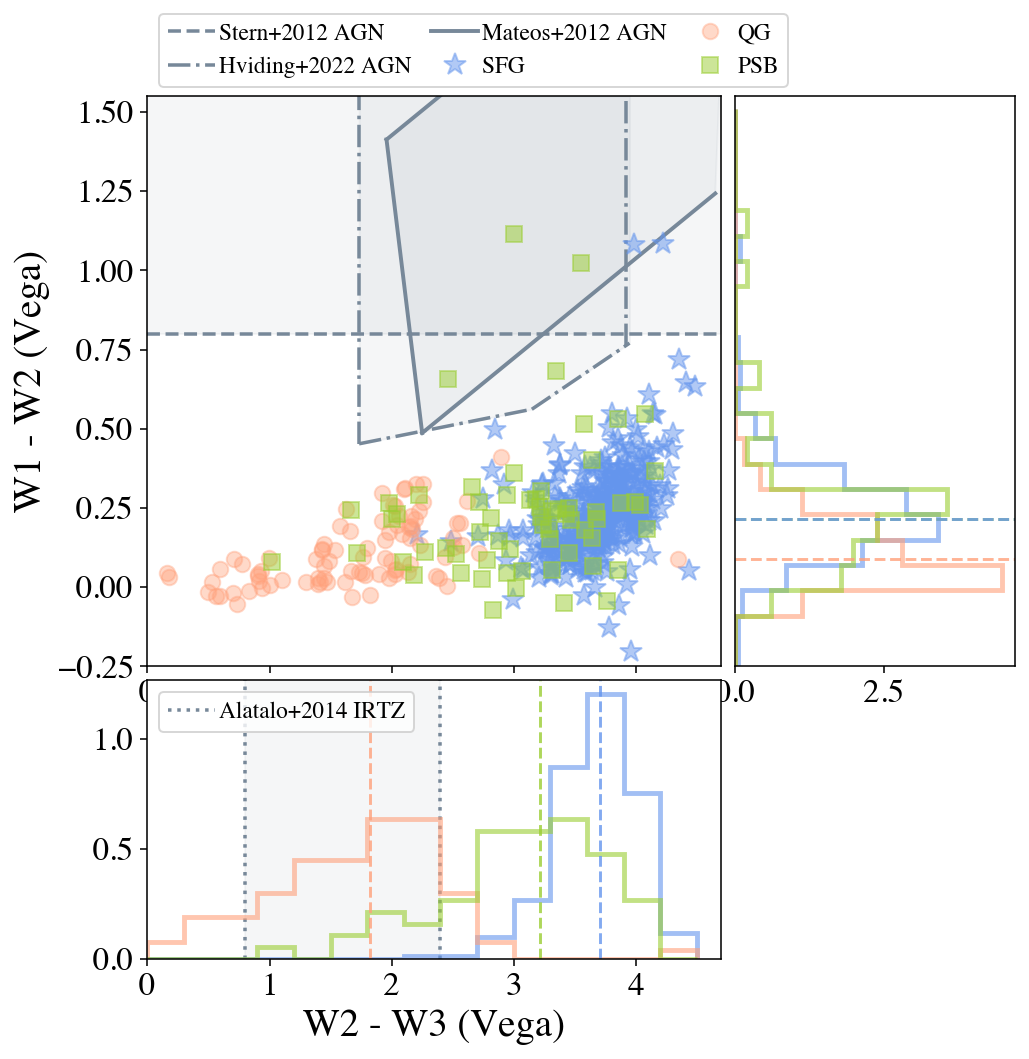}
\includegraphics[width=\columnwidth]{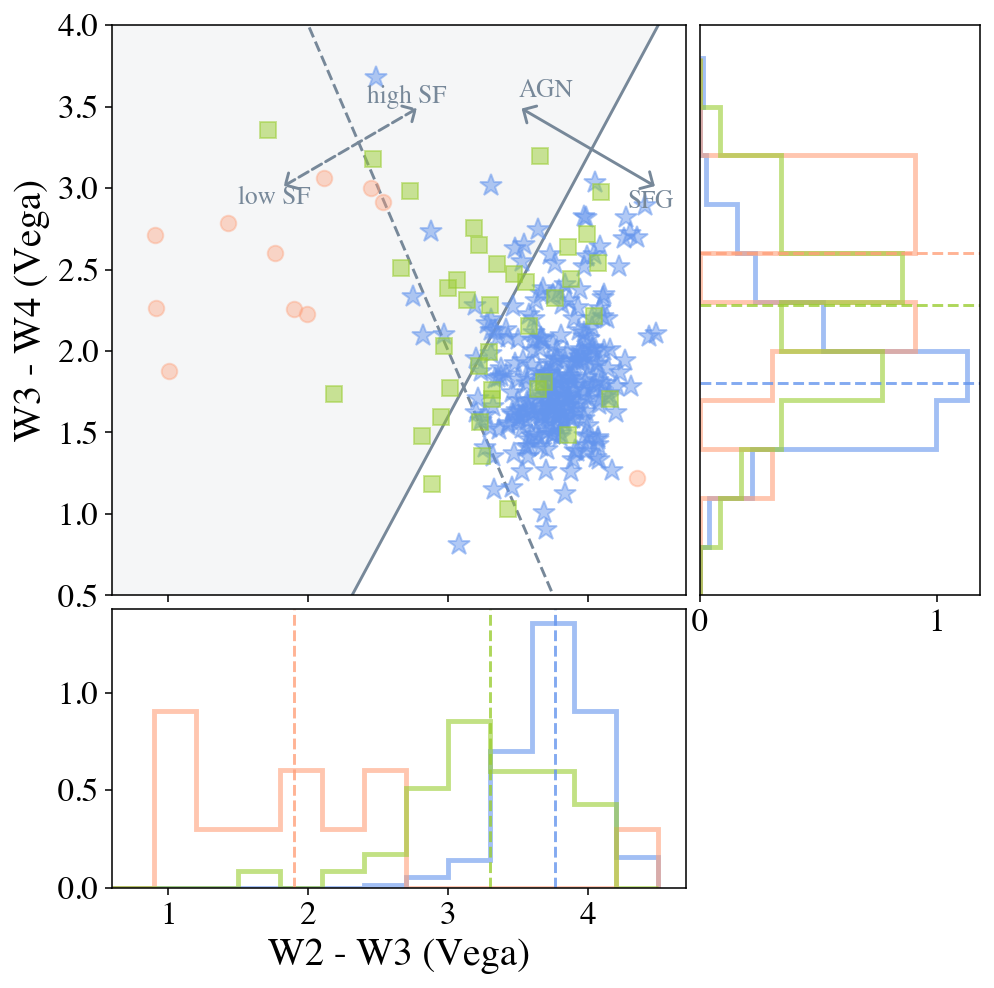}\\
\caption{\textbf{Left:} MIR color-color diagram of W1-W2 vs W2-W3. We overplot the AGN selection criteria from \citet{Stern_2012,Mateos_2012,Hviding_2022} in the main panel with different line styles and shading. In the bottom panel (W2-W3 histogram) we overplot the infrared transition zone (IRTZ) from \citet{Alatalo_2014_irtz}. \textbf{Right:} MIR color-color diagram of W3-W4 vs W2-W3. We overplot the diagnostic lines from \citet{Coziol_2015}. The AGN region is also shaded. On both panels only galaxies with SNR $>$ 3 in all relevant bands are plotted.
\label{fig:mir_color}}
\end{figure*}

MIR color has been used for AGN selection because emission from dust heated by AGN is expected to produce distinct colors from dust heated by stars \citep[e.g.,][]{Jarrett_2011,Stern_2012,Mateos_2012,Assef_2013}. In Figure \ref{fig:mir_color} we show the MIR color-color plots in both W1--W2 vs W2--W3 and W3--W4 vs W2--W3. On the W1--W2 vs W2--W3 panel, we overplot the AGN selection criteria from \citet{Stern_2012,Mateos_2012,Hviding_2022} and the infrared transition zone (IRTZ) from \citet{Alatalo_2014_irtz}. On the W3--W4 vs W2--W3 panel, we overplot the diagnostics from \citet{Coziol_2015}. AGN regions in both panels are shaded. While traditional MIR AGN color selection criteria \citep[e.g.,][]{Stern_2012,Mateos_2012} tend to select only the brightest ones, both \citet{Hviding_2022} and \citet{Coziol_2015} have accounted for low-luminosity AGN. Though the MIR AGN selections have reliability and completeness as high as $\sim$90\% and $\sim$75\% \citep[e.g.,][]{Stern_2012,Assef_2013,Assef_2018}, they could still miss some AGN in our galaxies. Since the same colors and criteria are used across samples, this effect does not bias our comparisons.

On the left panel of Figure \ref{fig:mir_color}, SFGs and PSBs show almost identical W1--W2 colors that appear slightly elevated compared to those of the QGs. The W2--W3 colors of the three galaxy populations show different distributions, with PSBs displaying intermediate colors between SFGs and QGs. Similar trends were observed by \citet{Alatalo_2017} within their samples of PSBs, early-, and late-type galaxies. However, the majority of our PSBs do not lie within the IRTZ proposed by \citet{Alatalo_2014_irtz,Alatalo_2017} as a potential criterion for identifying transitioning galaxies. This discrepancy likely arises from the differences in sample selection, as \citet{Alatalo_2017} selected comparison samples of early- and late-type galaxies based on morphology, and used the PSB sample from \citet{Goto_2007} which are known to be at a later stage of quenching \citep{French_2018}. While our PSB sample includes a small number of objects from \citet{Goto_2007}, it mostly consists of PSBs selected through other criteria that do not specifically exclude emission lines, potentially capturing younger PSBs closer to their starburst phase. Combining the three AGN selection criteria \citep{Stern_2012,Mateos_2012,Hviding_2022} plotted on this panel, 2 SFGs, 4 PSBs, and no QGs are selected as AGN based on their MIR colors. The AGN incidence rates for SFGs and PSBs are 0.6$^{+0.8}_{-0.4}$\% and 6.3$^{+4.8}_{-3.2}$\%\footnote{Uncertainties on the fractions in this section are Possion uncertainties following the same descriptions as in Table \ref{tab:det_rate}.}, respectively. Despite relatively large uncertainties, the AGN incidence rate in PSBs is at least 2 times higher than that in SFGs. Among the 2 SFGs and 4 PSBs identified as MIR AGN on this panel, none are detected in X-ray and 1 in each sample is detected in radio. Although the MIR AGN PSB and MIR AGN SFG both have L$_{\mathrm{1.4GHz,obs}}$/L$_{\mathrm{1.4GHz,SFR}}\sim1$, it is known that AGN selected in different wavelengths are not necessarily the same and no single waveband provides a complete and reliable selection of AGN \citep[e.g.,][]{Padovani_2017,Hickox_2018,Lyu_2022}. Detailed studies on individual objects are outside the scope of this work, and the detections are too few to allow statistically robust conclusions here.



On the right panel of Figure \ref{fig:mir_color}, SFGs and QGs are well separated by the demarcation line for low and high SF, although most QGs are not significantly detected in W3 and W4. PSBs again overlap with both SFGs and QGs, with more (74$^{+14}_{-14}$\% of PSBs) on the high SF side. This distribution likely reflects a combination of factors: our PSBs are at an early stage of quenching with some remaining SF activity, and MIR emission is sensitive to SF over a long timescale. 46$^{+10}_{-10}$\% of PSBs fall on the AGN side of the \citet{Coziol_2015} criterion and this AGN incidence rate is again at least 2 times higher than that of SFGs (5.8$^{+1.5}_{-1.5}$\%). All but one QGs are selected as AGN, although the low detection rates of QGs in W3 and W4 prevent us from making further statistical interpretations. Similar to the findings above, some MIR AGN identified by this method are also detected in radio and X-ray. Among the 18 MIR AGN PSBs, 2 are detected in X-ray and 3 are detected in radio (including one detected in both). None of the MIR AGN SFGs are detected in X-ray and 1 is detected in radio, while the MIR AGN QGs are undetected in both bands. The MIR AGN PSBs have L$_{\mathrm{1.4GHz,obs}}$/L$_{\mathrm{1.4GHz,SFR}}\sim4-25$ and L$_{\mathrm{0.2-2.3keV,obs}}$/L$_{\mathrm{0.2-2.3keV,XRB}}\sim0.3-2$ (see Section \ref{sec: agn in xray}). The MIR AGN SFG shows L$_{\mathrm{1.4GHz,obs}}$/L$_{\mathrm{1.4GHz,SFR}}\sim216$.

While AGB stars might potentially enhance the W3 luminosity of PSBs (Section \ref{sec:average mir}), this contribution will not shift PSBs to the shaded AGN regions in Figure \ref{fig:mir_color} and lead to the observed higher AGN incidence rate. The finding that PSBs exhibit a higher MIR AGN incidence rate than SFGs may be influenced by sample selection, as BPT AGN were excluded from the SFG sample but not from the PSBs. To assess this effect, we test the MIR AGN incidence in alternative samples of SFGs and QGs that includes BPT AGN (Appendix \ref{appendix new sample}). PSBs still show a higher MIR AGN fraction compared to this broader SFG sample, indicating that our results are not driven by sample selection biases.


The bolometric luminosity of AGN could be estimated from MIR luminosity assuming a bolometric correction factor. We apply the relation $L_{bol} = 10^{0.718}L_{3.4 \mu m}$ from \citet{Kim_2023} to the galaxies selected as MIR AGN based on the above color criteria. The resulting AGN bolometric luminosity values for the three samples are comparable and all are less than 9$\times 10^{44}$ erg/s. In contrast, the majority of $z<3$ WISE-selected AGN in \citet{Barrows_2021} have $L_{bol}>10^{45}$ erg/s. \citet{Kim_2023} derived this bolometric correction from SDSS quasars at $0.06<z<0.42$, a redshift range comparable to that of our samples. Although bolometric corrections could have relatively large uncertainties, they provide informative first-order estimations when complete SED modeling are not available. Our approximate estimation of $L_{bol}$ here should be regarded as upper limits, as the WISE W1 luminosity includes emission from both AGN in the nucleus and SF within the galactic disk. Thus although MIR-AGN appear to be more prevalent in PSBs, they have low-luminosity and their role in quenching remains inconclusive.

\subsection{AGN vs. SF in X-ray} \label{sec: agn in xray}

\begin{figure}
\centering
\includegraphics[width=\columnwidth]{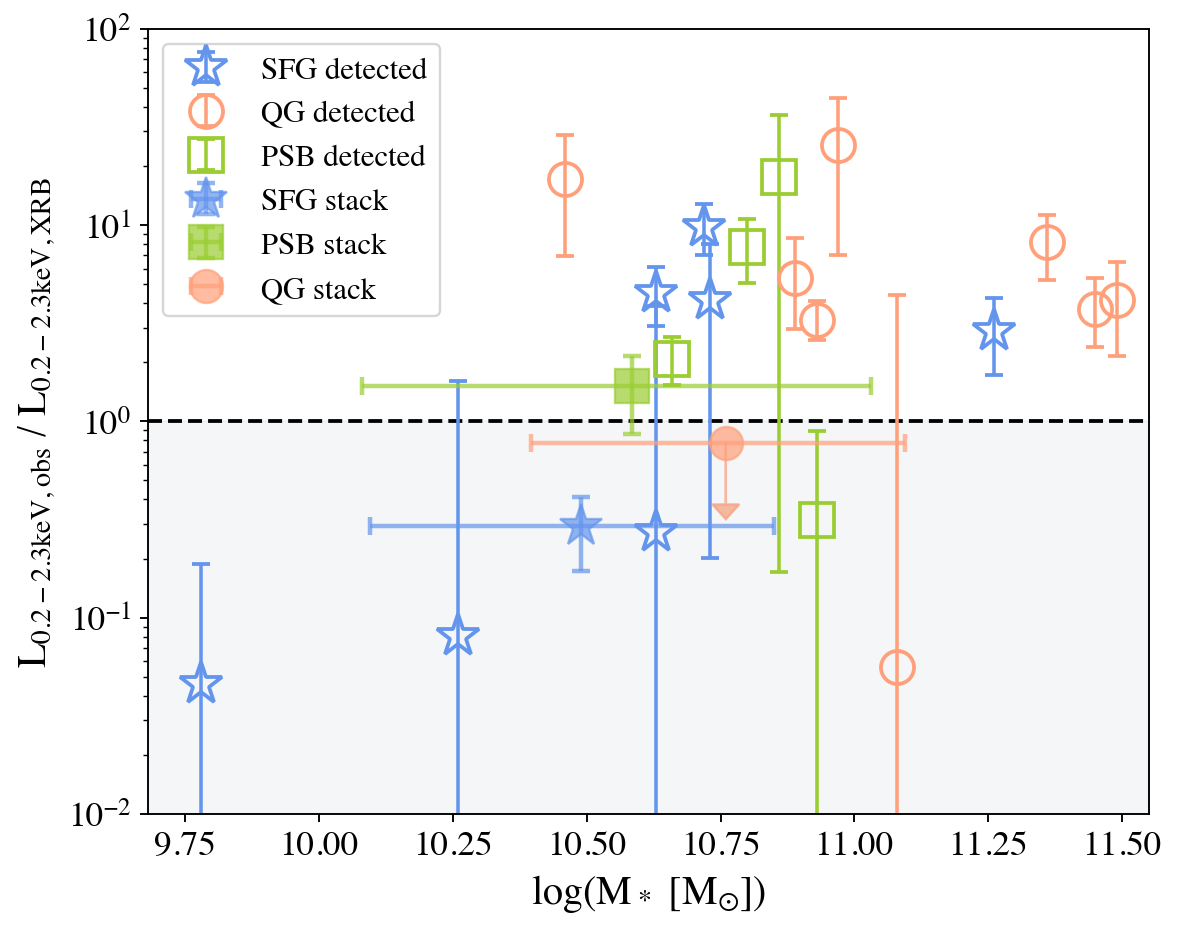}
\caption{Ratio of the observed X-ray luminosity and the expected X-ray luminosity from XRBs plotted against stellar mass. A ratio of L$_{\mathrm{0.2-2.3keV,obs}}$/L$_{\mathrm{0.2-2.3keV,XRB}}$ $<$ 1 indicates that the observed X-ray emission could be entirely coming from XRBs. Open points represent individually detected sources in the eFEDS catalog and filled points represent measurements from the stacked images (2$\sigma$ upper limit for the QG stack). The PSB stack exhibits L$_{\mathrm{0.2-2.3keV,obs}}$/L$_{\mathrm{0.2-2.3keV,XRB}}$ slightly above 1, enhanced relative to the SFG and QG stacks. The vertical errors are propagated from uncertainties in L$_{\mathrm{0.2-2.3keV,obs}}$ from the \citet{Liu_2022} catalog for individually detected sources, and from Section \ref{sec:xray stacking} for the stacks. See Section \ref{sec: agn in xray} for more discussions on uncertainties. Most of galaxies do not show significantly enhanced L$_{\mathrm{0.2-2.3keV,obs}}$/L$_{\mathrm{0.2-2.3keV,XRB}}$.}
\label{fig:l_int_l_xrb}
\end{figure}

We further compare the observed X-ray luminosity to that expected from SF to quantify the AGN level across our three samples. X-ray emission from SF arises from X-ray binaries (XRBs), which correlates with SFR and M$_*$. We estimate the X-ray luminosity from XRBs using the empirical scaling relation in \citet{Riccio_2023}, which is based on eFEDS data:
\begin{align}
    L_{\mathrm{0.2-2.3keV,XRB}} = 10^{29.25} M_* + 10^{39.95} \mathrm{SFR}.
\end{align}
We include both the low-mass and high-mass XRB terms (proportional to M$_*$ and SFR, respectively) for SFGs and PSBs, but only the low-mass XRB term for QGs due to their negligible SFR. We plot the ratio L$_{\mathrm{0.2-2.3keV,obs}}$/L$_{\mathrm{0.2-2.3keV,XRB}}$ for both individual detections and stacks in Figure \ref{fig:l_int_l_xrb}. A ratio of L$_{\mathrm{0.2-2.3keV,obs}}$/L$_{\mathrm{0.2-2.3keV,XRB}}<$ 1 indicates that XRBs (associated with SF activity) could explain all observed X-ray emission in the galaxy. While the hot halo gas in massive QGs also emits in X-ray, this emission is unlikely to be prominent in our QG stack given the faint signal from stacking $\sim$300 QGs (Section \ref{sec:xray stacking}). 

As illustrated in Figure \ref{fig:l_int_l_xrb}, individually detected galaxies in all three samples show a variety of L$_{\mathrm{0.2-2.3keV,obs}}$/L$_{\mathrm{0.2-2.3keV,XRB}}$ values. While some galaxies are consistent with the scenario where SF could explain all X-ray emission, some show L$_{\mathrm{0.2-2.3keV,obs}}$/L$_{\mathrm{0.2-2.3keV,XRB}}>1$, indicating additional X-ray emitters such as AGN. While the majority of individually detected QGs and PSBs exhibit L$_{\mathrm{0.2-2.3keV,obs}}$/L$_{\mathrm{0.2-2.3keV,XRB}}>1$, the overall range of this ratio is comparable across all three samples. Given the limited number of X-ray detections, we cannot statistically determine whether any one sample systematically exhibits higher or lower L$_{\mathrm{0.2-2.3keV,obs}}$/L$_{\mathrm{0.2-2.3keV,XRB}}$ than the others. This result is consistent with Figure \ref{fig:lx_ms_sfr}.

For the stacked measurement which represents the majority of galaxies in each sample, the PSB stack exhibits L$_{\mathrm{0.2-2.3keV,obs}}$/L$_{\mathrm{0.2-2.3keV,XRB}}$ $=1.5\pm0.6$, which is slightly elevated although still consistent with XRBs producing all X-ray emission. L$_{\mathrm{0.2-2.3keV,obs}}$/L$_{\mathrm{0.2-2.3keV,XRB}}$ for the SFG ($0.3\pm0.12$) and QG (0.78, upper limit) stacks are below 1. The scatter in the empirical XRB luminosity relations is typically $\sim$0.2--0.4 dex \citep[e.g.,][]{Lehmer_2010,Lehmer_2016}, so only sources with L$_{\mathrm{obs}}$/L$_{\mathrm{XRB}}$ greater than 3--5 are typically regarded as having potential AGN contribution \citep[e.g.,][]{Ito_2022}. While L$_{\mathrm{0.2-2.3keV,obs}}$/L$_{\mathrm{0.2-2.3keV,XRB}}$ for our stack of low-$z$ PSBs is not significantly elevated (e.g., $\gtrsim$ 5, as seen for QGs in \citealt{Ito_2022} at $z>1$), it is noticeably higher relative to those for the SFG and QG stacks due to the relatively larger L$_{\mathrm{0.2-2.3keV,obs}}$ and smaller SFR in the PSB stack. This observation suggests different X-ray properties between PSBs and SFGs/QGs and the potential presence of additional X-ray-emitting sources beyond XRBs (e.g., AGN) in PSBs. 

As discussed at the beginning of Section \ref{sec: multwavelength agn}, optical-MIR-based SFRs, and therefore derived L$_{\mathrm{0.2-2.3keV,XRB}}$, trace SF over a longer timescale and could be overestimated for PSBs. Thus L$_{\mathrm{0.2-2.3keV,obs}}$/L$_{\mathrm{0.2-2.3keV,XRB}}$ for PSBs could be higher than what is plotted in Figure \ref{fig:l_int_l_xrb}. PSBs have also been found in the literature to be relatively dust-rich contrary to the expectation that they should contain little dust \citep[e.g.,][]{Smercina_2018,Li_2019}, which could obscure their X-ray emission. However, we do not have enough information to constrain the intrinsic obscuring column density in the stacks. The small number of individually detected galaxies and the large uncertainties ($\sim$0.9 dex) on N$_{\rm H}$ from \citet{Liu_2022} for them prevent us from characterizing the representative obscuration levels in our samples. Although L$_{\mathrm{0.2-2.3keV,obs}}$/L$_{\mathrm{0.2-2.3keV,XRB}}$ for PSBs is consistent with 1, our analyses do not rule out the possibility of obscured and/or weak AGN in the PSB population. Our X-ray findings are consistent with \citet{Almaini_2025}, who reported only weak AGN activity in PSBs at $1<z<3$, with X-ray luminosities comparable to those of older passive galaxies and no evidence of elevated AGN activity.

\section{Discussion} \label{sec:discussion}

In this section, we discuss the potential connections and associated uncertainties between AGN and quenching in light of our results.

Our findings (Section \ref{sec:agn in radio}, \ref{sec: agn in mir}, \ref{sec: agn in xray}) do not reveal prominent AGN activity in PSBs from radio, MIR, and X-ray emission. While PSBs overall do not exhibit significantly enhanced AGN emission compared to SFGs and QGs, they do show an elevated MIR AGN incidence rate and a slightly higher stacked L$_{\mathrm{0.2-2.3keV,obs}}$/L$_{\mathrm{0.2-2.3keV,XRB}}$ ratio. While the elevated MIR AGN fraction in PSBs could arise from the exclusion of optical AGN in the SFG and QG samples, an alternative sample set retaining optical AGN in the SFGs and QGs yield similar results (Appendix \ref{appendix new sample}). The possibility that PSBs host an overabundance of low-luminosity, potentially obscured AGN relative to SFGs and QGs remains intriguing. Nevertheless, direct evidence of AGN driving SF quenching remains elusive.

One major challenge in establishing the causal link between AGN and quenching lies in the stochastic nature of the AGN duty cycle. As a galaxy moves from star-forming to quiescent, its central black hole could switch between active and inactive multiple times (see Figure 9 of \citealt{Almaini_2025} for a nice illustration). BH growth at high accretion rates could last $\sim$ 10$^5$ yrs \citep[e.g.,][]{King_2015,Schawinski_2015}. Similarly, \citet{French_2023} found evidence for fading AGN using spatially resolved data in five nearby PSBs and estimated the AGN duty cycle during the PSB phase to be $\sim$ 10$^5$ yrs. \citet{French_2023} also estimated that AGN spend only 5.3\% of this duty cycle in the luminous phase (see also \citealt{Almaini_2025}) and proposed that the duration of the duty cycle may explain why so few luminous AGNs have been observed during the PSB phase. AGN activity may also peak prior to quenching, fueled by the same gas reservoir that supported earlier SF, rather than acting as the primary driver of quenching. Galaxy mergers could significantly affect the gas distribution in galaxies and are associated with both AGN and SF/SB. PSBs are frequently observed to harbor merger signatures such as disturbed morphologies \citep[e.g.,][]{Pawlik_2016,Chen_2019,Sazonova_2021,Ellison_2022,Verrico_2023,Ellison_2024_post-merger_quenching}. With neural network predicted post-merger times, \citet{Ellison_2024_post-merger_agn} found that AGN and SB activities peak contemporaneously (with a time difference $\lesssim$150 Myr), immediately after coalescence. In contrast, the fraction of PSBs in post-mergers peaks between 200--500 Myr after coalescence \citep{Ellison_2024_post-merger_quenching}. Similarly, \citet{Greene_2020} found that at $z \sim$ 0.7, the decrease in AGN fraction from young to old PSBs coincides with a decrease in the molecular gas fraction, suggesting AGN in those PSBs are more likely associated with fueling rather than feedback. 

Another puzzle is how long it takes for AGN to quench SF. Studies based on both simulations and observations have found that whether a galaxy is quenched most closely associates with its black hole mass (M$_{\rm BH}$, e.g., \citealt{Terrazas_2016,Piotrowska_2022,Bluck_2023}), rather than its observed instantaneous AGN luminosity (related to $\Dot{\rm M}_{\rm BH}$). These works suggest that it is the integrated power output from the AGN over a galaxy's past lifetime that quenches its SF. Spatially resolved studies \citep[e.g.,][]{Lammers_2023} also reveal that while Gyr-timescale AGN feedback can suppress central SF, it may be inefficient in driving galaxy-wide quenching, leaving the galaxy in the green valley (see also \citealt{luo2022}). If more violent AGN episodes occurred earlier and have been contributing to the cumulative impact of feedback in quenching, the low-level AGN activity suggested by our X-ray stacking analysis, along with the elevated MIR AGN incidence in PSBs, could be the afterglow of those past events.


Our results are consistent with both AGN quenching (AGN is contributing to SF suppression) and AGN fueling (AGN in PSBs are mere consequences of past SB events which could be triggered by mergers). A combination of both AGN quenching and fueling is also plausible. The observation that QGs exhibit significantly higher L$_{\rm obs}$/L$_{\rm SFR}$ ratios than PSBs in the radio but comparable (even lower in the stack) ratios in X-ray also hints at possible different dominant feedback channels between these evolutionary stages (e.g., the radiative and kinetic mode feedback). However, a detailed discussion on different AGN feedback mechanisms is outside the scope of this work and the limited sample sizes prevent us from making further conclusions. We emphasize that the AGN strengths discussed here are relative within our samples, which do not contain a significant fraction of strong AGN (such as radio-loud or X-ray-bright ones), and are focused on the low-redshift universe and the faint end of the AGN luminosity function. 

Although this study does not definitively distinguish between AGN fueling and quenching during the PSB phase, the absence of bright AGN and the lack of excess AGN emission relative to SFGs and QGs suggest that the role of AGN in quenching at low redshift is more subtle than blowing out gas \citep[e.g.,][]{Hopkins_2006,Somerville_2008,Costa_2020} -- the feedback is likely more ``preventive" than ``ejective". Future work with larger samples (e.g., utilizing the eROSITA all-sky survey) will improve the statistical power of this study, allowing stacking in different bins of galaxy properties (e.g., M$_*$ and post-burst age) for a more comprehensive understanding of AGN incidence in quenching galaxies. Follow-up observations of PSBs in harder X-ray bands (e.g., above 5 keV) will be particularly valuable for detecting obscured and low-luminosity AGN, as lower-energy X-ray photons are more susceptible to absorption. In addition, analyses on SFGs, PSBs, and QGs at higher redshifts where the quenching of most local massive galaxies happened are essential for developing a comprehensive understanding of quenching across cosmic history (e.g., \citealt{Almaini_2025}, Patil et al., in prep).

\section{Conclusions} \label{sec: conclusions}

While multiwavelength studies of quenching galaxies offer valuable insights into this critical transitional stage of galaxy evolution, studies to date have been limited in number statistics. In this work, we investigate AGN activity in nearby ($z<$ 0.2) SFGs, PSBs, and QGs combining surveys in X-ray, MIR, and radio. Our goal is to better characterize AGN strength across different galaxy populations (evolutionary stages) and assess whether AGN activity is elevated during the quenching phase. Our primary conclusions are:

\begin{enumerate}
    \item The majority of our galaxies are not individually detected in radio and X-ray surveys (Table \ref{tab:det_rate}). While PSBs show $\sim$ 2 times higher detection rates than SFGs and QGs in radio and X-ray, the detection rates for all three galaxy populations may still be consistent within statistical uncertainties. Most individually detected galaxies do not fall in the radio-loud or X-ray-bright regime and we perform image stacking for the non-detections to recover the signal representative of each galaxy sample. We measure SNR $>$ 2 signals in the radio stacks for SFGs, PSBs, and QGs. We measure SNR $>$ 2 signals in the X-ray stacks for SFGs, PSBs, and obtain a 2$\sigma$ upper limit in the X-ray stack for QGs.

    \item We do not observe elevated AGN activity in PSBs in radio emission. By comparing observed the radio luminosity to that expected from SF (L$_{\mathrm{1.4GHz,obs}}$/L$_{\mathrm{1.4GHz,SFR}}$, Figure \ref{fig:lrad_ms}), QGs (individual detections and stack) host relatively stronger AGN than both SFGs and PSBs. Individually detected SFGs and PSBs exhibit comparable L$_{\mathrm{1.4GHz,obs}}$/L$_{\mathrm{1.4GHz,SFR}}$ values around 5. Given the sensitivity of radio emission to longer timescale SF and the bursty SF histories of PSBs, this result is consistent with a scenario in which all radio emission in these PSBs originates from SF. Stacked radio luminosities for both SFGs and PSBs align with expectations from SF, indicating that radio-undetected PSBs do not harbor weak radio AGN activity.  
    
    

    \item PSBs exhibit a elevated MIR AGN incidence rate. Based on the MIR color diagnostics (Figure \ref{fig:mir_color}), the AGN incidence rate in PSBs is at least twice that in SFGs on both W1-W2 vs W2-W3 and W3-W4 vs W2-W3 plots. None of the QGs lie in the AGN regions on the W1-W2 vs W2-W3 plot, and few are significantly detected in W3 and W4 bands. Although the elevated MIR AGN fraction in PSBs could arise from the exclusion of optical AGN in the SFG and QG samples, we test this with an alternative sample set retaining optical AGN in the SFGs and QGs and obtain similar results (Figure \ref{fig:mir_color_new_sfg}). However, the MIR AGN fractions for both sample sets are subject to small number statistics. The bolometric luminosities estimated for MIR-selected AGN in all three samples are comparable and small ($<9\times 10^{44}$ erg/s), and their role in quenching remains inconclusive. PSBs show an intermediate composite MIR SED between SFGs and QGs from the four WISE bands. The gradual changes in MIR luminosity and colors from SFGs to QGs could reflect the evolving trend of a decrease (increase) in the SF (AGN) contribution to the total galaxy MIR emission.
    
    


    \item We do not observe elevated AGN activity in PSBs in X-ray emission. Comparing observed X-ray luminosity to that expected from SF (L$_{\mathrm{0.2-2.3keV,obs}}$/L$_{\mathrm{0.2-2.3keV,XRB}}$, Figure \ref{fig:l_int_l_xrb}), individually detected SFGs, PSBs, and QGs show significant overlap with some galaxies exhibiting values exceeding 1 indicating non-SF X-ray emission likely from AGN. While L$_{\mathrm{0.2-2.3keV,obs}}$/L$_{\mathrm{0.2-2.3keV,XRB}}$ for the stacks are consistent with an SF origin of X-ray emissions in all three galaxy samples, the PSB stack shows a ratio $\gtrsim$ 1, noticeably higher than those of the SFG and QG stacks. Considering the uncertainties in estimating L$_{\mathrm{XRB}}$ for PSBs, their potential for greater dust obscuration, and their elevated MIR AGN fraction, this result does not rule out the presence of obscured, low-luminosity AGN within the PSB population.

\end{enumerate}

Our multiwavelength investigation does not reveal prominent AGN activity in nearby PSBs. These PSBs overall do not exhibit significantly enhanced AGN emission compared to mass- and redshift-matched SFGs and QGs, aside from tentative evidence in the MIR and X-ray. If AGN are prevalent in nearby PSBs, they are likely weak or obscured. Those AGN could be fueled by the same gas reservoir that sustained the preceding starburst and may or may not contribute to subsequent quenching, as seen in literature studies. AGN's role in quenching at low redshift is more subtle than violently removing the gas -- the feedback is likely more ``preventive" than ``ejective". Follow-up studies utilizing harder X-ray observations will help identify heavily obscured AGN in PSBs. A closer view of the multiphase gas properties will also shed light on why PSBs are not efficiently forming stars.



\section*{Acknowledgements}
We thank the anonymous reviewer for the thorough comments that improve the clarity of the manuscript. We thank Dustin Lang for helping to obtain the unWISE catalog for the eFEDS field. P.P. and M.S. gratefully acknowledge support from the NASA Astrophysics Data Analysis Program (ADAP) under grant 80NSSC23K0495. J.O. acknowledges support from the Space Telescope Science Institute Director's Discretionary Research Fund grants D0101.90296 and D0101.90311. Y.L. acknowledges support from the Space Telescope Science Institute Director's Discretionary Research Fund grant D0101.90281. J.S. acknowledges support from the National Science Foundation under NSF-AAG No. 2407954. O.A. acknowledges the support from STFC grant ST/X006581/1. This research made use of \texttt{Photutils}, an Astropy package for detection and photometry of astronomical sources \citep{photutils}. Funding for the SDSS and SDSS-II has been provided by the Alfred P. Sloan Foundation, the Participating Institutions, the National Science Foundation, the U.S. Department of Energy, the National Aeronautics and Space Administration, the Japanese Monbukagakusho, the Max Planck Society, and the Higher Education Funding Council for England. The SDSS Web Site is \url{http://www.sdss.org/}. The SDSS is managed by the Astrophysical Research Consortium for the Participating Institutions. The Participating Institutions are the American Museum of Natural History, Astrophysical Institute Potsdam, University of Basel, University of Cambridge, Case Western Reserve University, University of Chicago, Drexel University, Fermilab, the Institute for Advanced Study, the Japan Participation Group, Johns Hopkins University, the Joint Institute for Nuclear Astrophysics, the Kavli Institute for Particle Astrophysics and Cosmology, the Korean Scientist Group, the Chinese Academy of Sciences (LAMOST), Los Alamos National Laboratory, the Max-Planck-Institute for Astronomy (MPIA), the Max-Planck-Institute for Astrophysics (MPA), New Mexico State University, Ohio State University, University of Pittsburgh, University of Portsmouth, Princeton University, the United States Naval Observatory, and the University of Washington. This publication makes use of data products from the Wide-field Infrared Survey Explorer, which is a joint project of the University of California, Los Angeles, and the Jet Propulsion Laboratory/California Institute of Technology, funded by the National Aeronautics and Space Administration. This work is based on data from eROSITA, the soft X-ray instrument aboard SRG, a joint Russian-German science mission supported by the Russian Space Agency (Roskosmos), in the interests of the Russian Academy of Sciences represented by its Space Research Institute (IKI), and the Deutsches Zentrum f$\ddot{u}$r Luft- und Raumfahrt (DLR). The SRG spacecraft was built by Lavochkin Association (NPOL) and its subcontractors, and is operated by NPOL with support from the Max Planck Institute for Extraterrestrial Physics (MPE). The development and construction of the eROSITA X-ray instrument was led by MPE, with contributions from the Dr. Karl Remeis Observatory Bamberg \& ECAP (FAU Erlangen-Nuernberg), the University of Hamburg Observatory, the Leibniz Institute for Astrophysics Potsdam (AIP), and the Institute for Astronomy and Astrophysics of the University of T$\ddot{u}$bingen, with the support of DLR and the Max Planck Society. The Argelander Institute for Astronomy of the University of Bonn and the Ludwig Maximilians Universit$\ddot{a}$t Munich also participated in the science preparation for eROSITA.


\facilities{eROSITA, WISE, VLA FIRST, SDSS}


\software{astropy \citep{2013A&A...558A..33A,2018AJ....156..123A}, astroquery \citep{astroquery}, SciServer \citep{sciserver}, photutils \citep{photutils}, WebPIMMS}

\appendix

\section{Comparison Sample Matching} \label{appendix control sample}

\begin{figure}[h]
\centering
\includegraphics[width=\columnwidth]{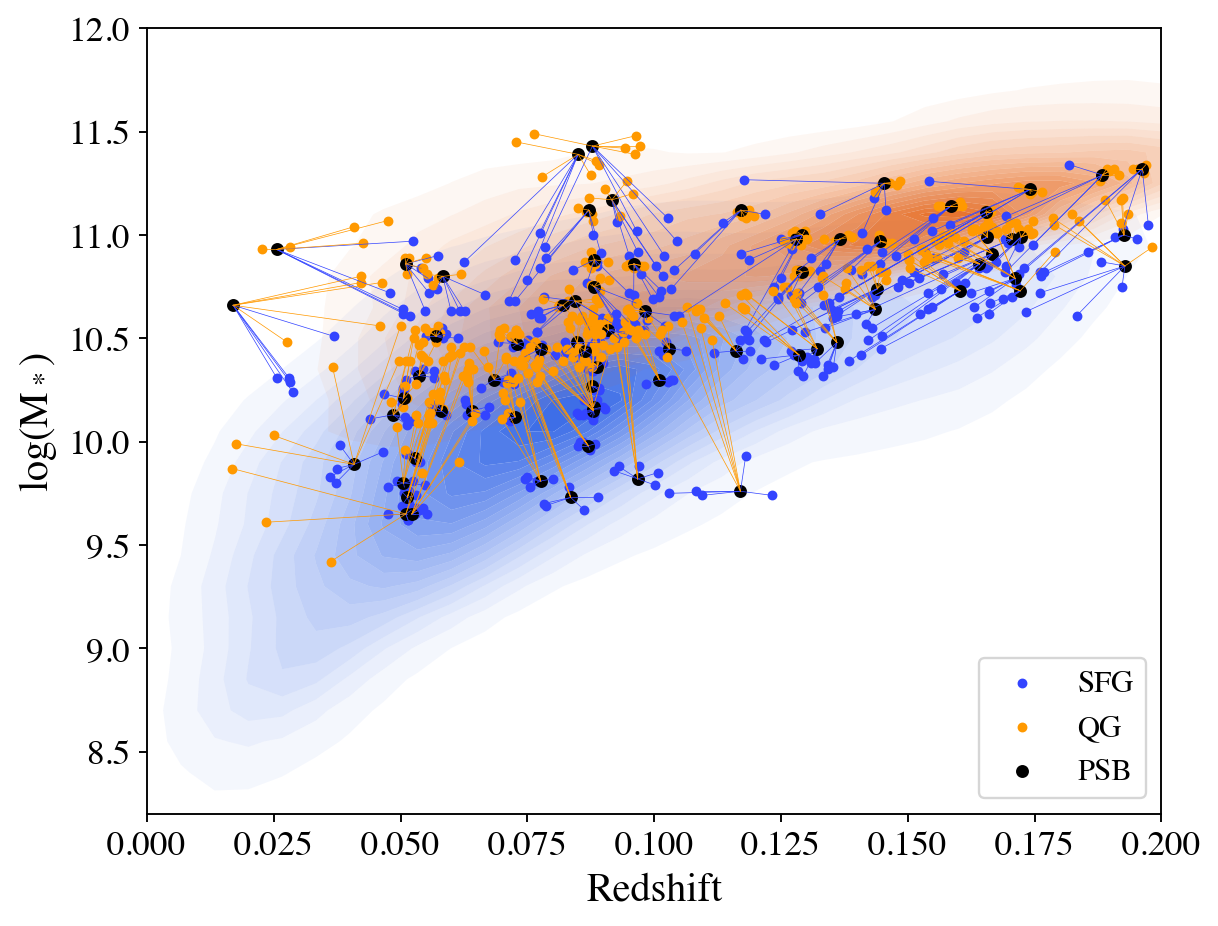}
\caption{The assignment scheme used to construct the SFG and QG comparison samples for PSBs. PSBs and their matched SFGs and QGs are connected by lines, and the color-shaded background shows the number density of the parent samples of SFGs (1136 galaxies) and QGs (1102 galaxies) in the eFEDS field.
\label{fig:sample_assigment}}
\end{figure}

Figure \ref{fig:sample_assigment} illustrates our comparison sample assignment strategy, which follows the Hungarian algorithm described in \citet{Sazonova_2021} to minimize the total differences in M$_*$ and $z$ across all matched pairs. A significant number of QGs are located near the lower mass boundary of the parent QG sample (appearing as the ``diagonal line" pattern), due to the scarcity of low-mass quiescent galaxies (M$_*$ $<$ 10$^{10}$ M$_{\odot}$) at $z<0.2$. This dearth of low-mass QGs has also been observed in \citet{Chang_2015} (e.g., their Figure 13) and reflects the general lack of low-mass QGs at $z<0.2$ in SDSS DR7. There also appears to be an aggregation of QGs around a vertical line at $z\sim0.1$ following the distribution of PSBs. It is possible that there is an overdensity of PSBs at $z\sim0.1$ but detailed investigations on environments of PSBs are outside the scope of this work. We note that these patterns are not biases in the matching process.

While individual matched pairs exhibit different $\Delta$M$_*$ and $\Delta z$, the SFG, PSB, QG samples are analyzed collectively in each category. As described in Section \ref{sec:control sample}, we discard matches with $\Delta$M$_*>0.4$ dex and $\Delta z$ is $\lesssim0.01$ for $\sim$90\% of the matches. We verified that imposing stricter criteria of $\Delta$M$_*<0.2$ dex and $\Delta z<0.005$ on the comparison samples does not alter our conclusions, although the reduced sample sizes increase statistical uncertainties in the stacking analysis. Therefore, our SFG and QG samples are sufficiently matched in M$_*$ and $z$ to ensure fair comparisons.

\section{Alternative Comparison Samples} \label{appendix new sample}

As described in Section \ref{sec:control sample}, we apply both H$\delta_A$ vs. Dn4000 and BPT criteria when constructing comparison samples of SFGs and QGs. While these clean baseline comparison samples facilitate clear characterization of the average multiwavelength properties for each galaxy population, minimizing the risk of AGN dominating the stacked signals, they may introduce biases when comparing AGN prevalence (e.g., in the MIR, Section \ref{sec: agn in mir}). Here we test whether the elevated MIR AGN incidence rate in PSBs found in Section \ref{sec: agn in mir} is a result of sample selection, i.e., whether the lower MIR AGN fraction in SFGs and the limited statistics for QGs are due to the prior exclusion of BPT (optical) AGN.

\begin{figure*}[h]
\centering
\includegraphics[width=\textwidth]{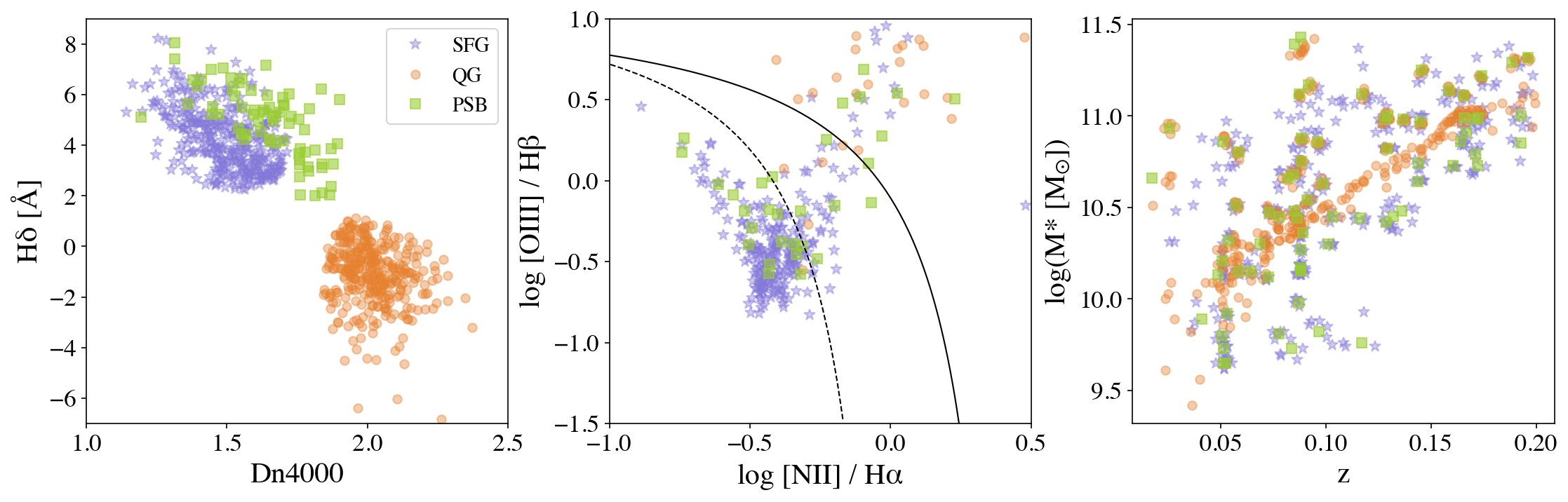}
\caption{Same as Figure \ref{fig:sample}, but for the alternative test comparison samples in Appendix \ref{appendix new sample}. \textbf{Left:} The distribution of SFGs, PSBs and QGs on the H$\delta_A$ vs. Dn4000 plot. \textbf{Middle:} [\ion{O}{3}]/H$\beta$ vs. [\ion{N}{2}]/H$\alpha$ BPT diagram. The dashed and solid lines are demarcation lines for SF and AGN ionization, from \citet{Kauffmann_2003} and \citet{Kewley_2001}, respectively. Only galaxies with SNR $\geqslant$ 3 for all four emission lines are shown. \textbf{Right:} Stellar mass vs. redshift for the PSB sample and mass- and redshift-matched comparison samples of SFGs and QGs. The vertical and diagonal aggregations of QGs in this panel are not biases introduced by the matching process. See Section \ref{sec:control sample} and Appendix \ref{appendix control sample} for more details.
\label{fig:sample_new_sfg}}
\end{figure*}

We construct alternative test samples of SFGs and QGs following the same procedures detailed in Section \ref{sec:control sample}, but without applying any BPT-based selection criteria. As before, we match each PSB 5 SFGs and 5 QGs in both M$_*$ and $z$, requiring a mass difference of less than 0.4 dex to ensure a close match. This results in a sample of 363 SFGs and 347 QGs for the 73 PSBs. The locations of these galaxies on diagnostic diagrams are shown in Figure \ref{fig:sample_new_sfg}. The BPT AGN fractions for these samples of SFGs, PSBs, and QGs are 5.5$^{+1.5}_{-1.5}$\%, 20$^{+12}_{-8}$\%, 90$^{+14}_{-14}$\%, after requiring SNR $\geq$ 3 for all BPT lines. Uncertainties on the fractions in this section are Possion uncertainties following the same descriptions as in Table \ref{tab:det_rate}. PSBs do have higher optical AGN fractions than SFGs, consistent with prior studies that suggest an overabundance of AGN in the PSB phase (Section \ref{sec:intro}). QGs, as expected, show predominantly AGN-like optical line emission as they have minimal SF by definition.

\begin{figure*}
\centering
\includegraphics[width=\columnwidth]{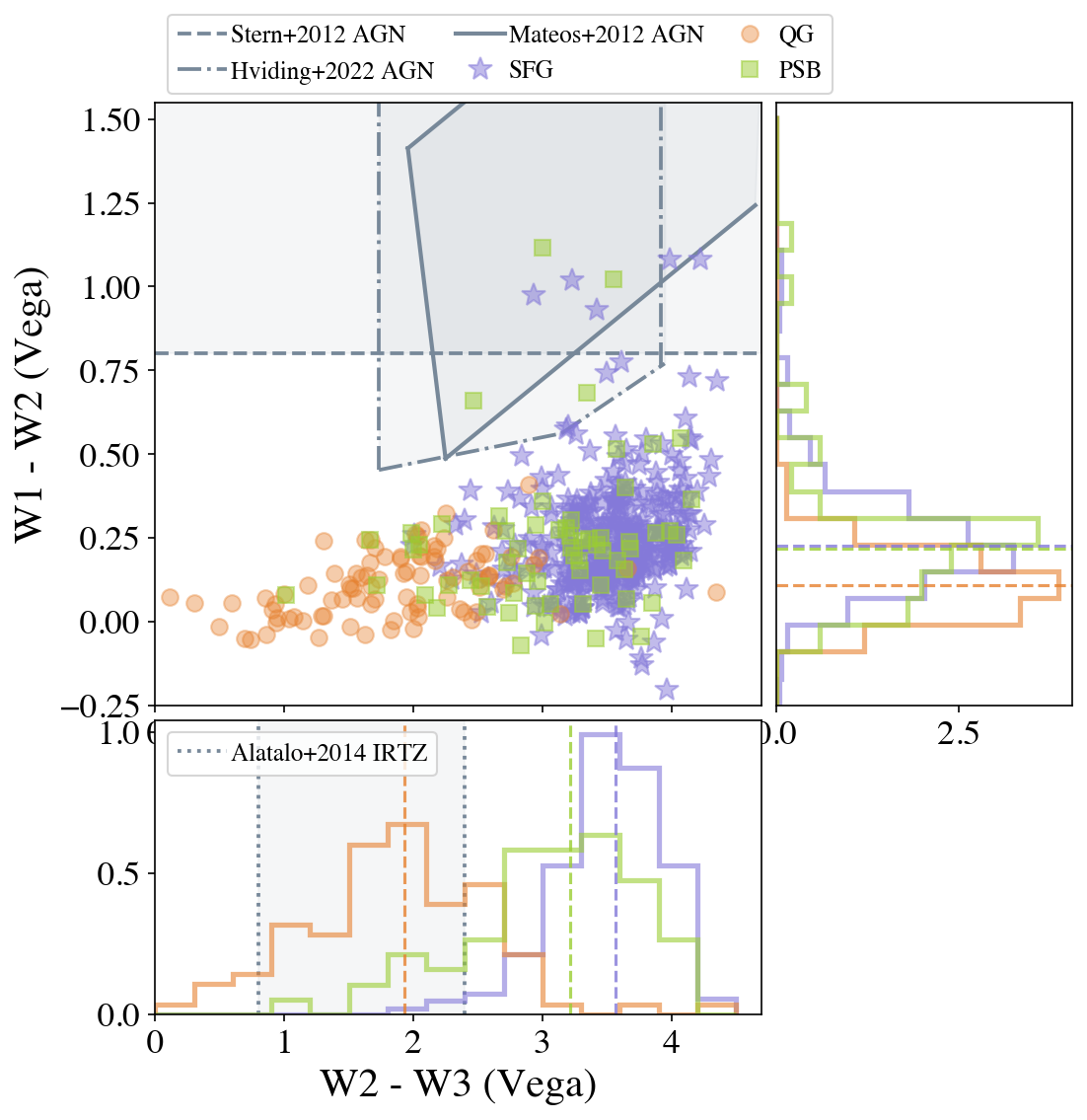}
\includegraphics[width=\columnwidth]{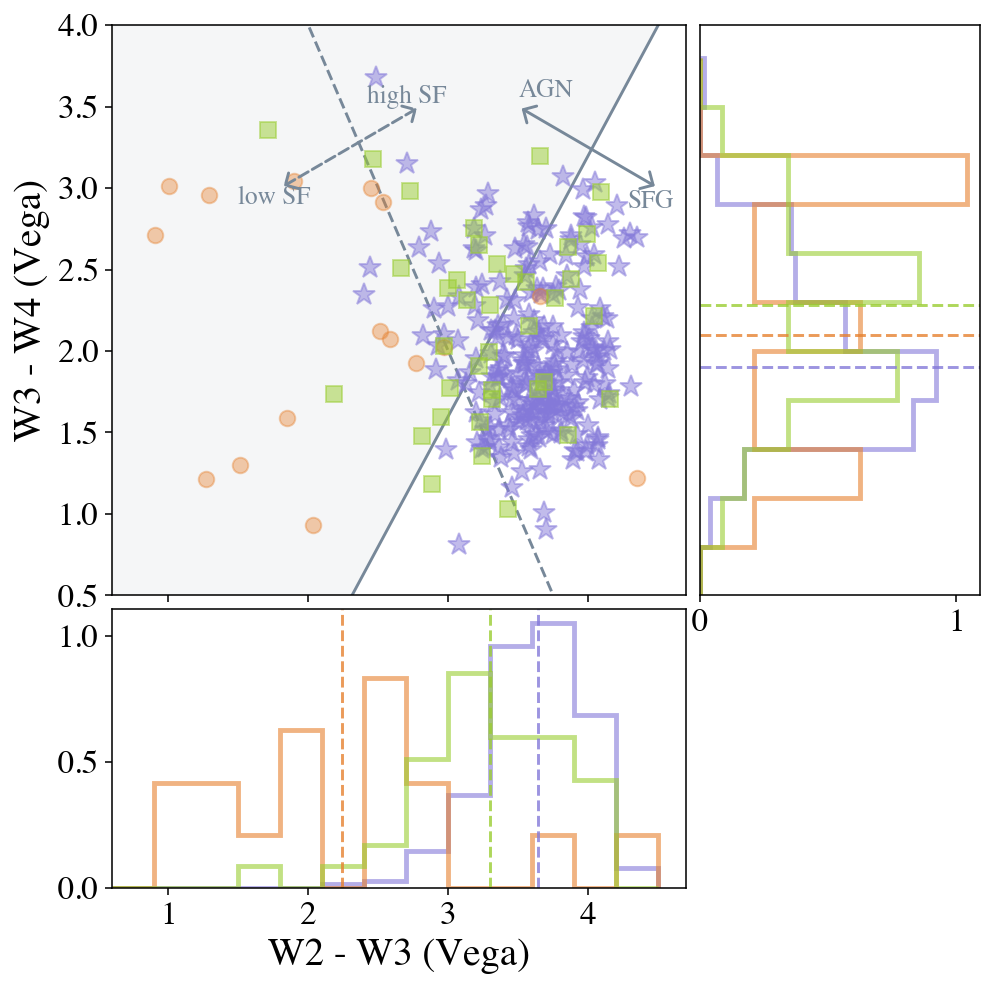}\\
\caption{Same as Figure \ref{fig:mir_color}, but for the alternative test comparison samples in Appendix \ref{appendix new sample}. \textbf{Left:} MIR color-color diagram of W1-W2 vs W2-W3. We overplot the AGN selection criteria from \citet{Stern_2012,Mateos_2012,Hviding_2022} in the main panel with different line styles and shading. In the bottom panel (W2-W3 histogram) we overplot the infrared transition zone (IRTZ) from \citet{Alatalo_2014_irtz}. \textbf{Right:} MIR color-color diagram of W3-W4 vs W2-W3. We overplot the diagnostic lines from \citet{Coziol_2015}. The AGN region is also shaded. On both panels only galaxies with SNR $>$ 3 in all relevant bands are plotted.
\label{fig:mir_color_new_sfg}}
\end{figure*}

We reproduce the MIR color-color diagrams from Section \ref{sec: agn in mir} using these test samples in Figure \ref{fig:mir_color_new_sfg}. The distributions of the three populations remain consistent with those shown in Figure \ref{fig:mir_color}. On the left panel (W1--W2 vs W2--W3), more SFGs now fall in the AGN selection region, though the fraction (1.9$^{+1.1}_{-0.7}$\%) is still lower than that of PSBs (6.3$^{+4.8}_{-3.2}$\%). No QGs are selected as AGN based on these colors. On the right panel (W3--W4 vs W2--W3), the MIR AGN fractions for SFGs and PSBs are 17$^{+2.6}_{-2.6}$\% and 46$^{+10}_{-10}$\%, respectively. QGs remain largely undetected in W3 and W4, limiting statistical interpretation. Overall, these results are consistent with those presented in Section \ref{sec: agn in mir} and confirm that the elevated MIR AGN incidence in PSBs is not driven by sample selection effects. In terms of radio and X-ray analyses, this selection will not affect our conclusions. If the presence of a few BPT AGN in PSBs is biasing the result, we would expect PSBs to show elevated AGN level compared to the clean baseline samples of SFGs/QGs that do not contain BPT AGN. In contrast, we do not observe elevated AGN activity in PSBs in radio and X-ray.

\bibliography{sample631,classics}{}
\bibliographystyle{aasjournal}

\end{document}